\documentclass[fleqn,usenatbib]{mnras}

\usepackage{newtxtext,newtxmath}

\usepackage[T1]{fontenc}
\usepackage{ae,aecompl}

\usepackage{graphicx}	
\usepackage{amsmath}	
\usepackage{amssymb}	
\usepackage[usenames,dvipsnames]{xcolor}
\usepackage{soul}

\newcommand{\lp}{LP\,40$-$365}
\newcommand{\jg}{\mbox{J1603$-$6613}}
\newcommand{\jb}{J1825$-$3757}
\newcommand{\jc}{J0905+2510}
\newcommand{\jd}{J1637+3631}
\newcommand{\dox}{J1240+6710}
\newcommand{\Teff}{\mbox{$T_\mathrm{eff}$}}
\newcommand{\kms}{\mbox{km\,s$^{-1}$}}

\newcommand{\masyr}{\mbox{mas\,yr$^{-1}$}}
\newcommand{\chisqr}{\mbox{$\chi^2_\nu$}}
\newcommand{\sdss}[3]{SDSS\,J#1$#2$#3}
\newcommand{\Ion}[2]{#1\,\textsc{#2}}

\usepackage[dvipsnames]{xcolor}
\usepackage{soul}

\defcitealias{raddi18a}{Paper~I}
\defcitealias{raddi18b}{Paper~II}

   \title[Partly burnt post-supernova runaway remnants]{Partly burnt runaway stellar remnants from peculiar thermonuclear supernovae}

\author[R. Raddi et al.]{R. Raddi,$^1$\thanks{E-mail: roberto.raddi@fau.de}
          M. A. Hollands,$^{2}$
          D. Koester,$^{3}$
          J. J. Hermes,$^{4,5}$
          B. T. G\"ansicke,$^{2}$
          U. Heber,$^{1}$
\newauthor{K. J. Shen,$^{6}$
          D. M. Townsley,$^{7}$
          A. F. Pala,$^{2,8}$
          J. S. Reding,$^{5}$
          O. F. Toloza,$^{2}$
          I. Pelisoli,$^{9}$}
\newauthor{S. Geier,$^{9}$
          N. P. Gentile Fusillo,$^{2}$
          U. Munari,$^{10}$
          J. Strader$^{11}$}
\\
$^{1}$Dr. Remeis-Sternwarte, Friedrich Alexander Universit\"at Erlangen-N\"urnberg, Sternwartstr. 7, 96049 Bamberg, Germany\\
$^{2}$University of Warwick, Department of Physics, Gibbet Hill Road, Coventry, CV4 7AL, United Kingdom\\
$^{3}$Universit\"at Kiel, Institut f\"ur Theoretische Physik und Astrophysik, 24098, Kiel, Germany\\
$^{4}$Department of Astronomy, Boston University, 725 Commonwealth Ave., Boston, MA 02215, US\\
$^{5}$University of North Carolina, Department of Physics and Astronomy, Chapel Hill, NC - 27599-3255, US\\
$^{6}$Department
of Astronomy and Theoretical Astrophysics Center, University of California, Berkeley, CA 94720, USA\\
$^{7}$University of Alabama, Department of Physics and Astronomy, Tuscaloosa, AL, USA\\
$^{8}$European Southern Observatory, Karl-Schwarzschild-Str. 2, 85748, Garching\\
$^{9}$Institut fur Physik und Astronomie, Universit\"at Potsdam, Haus 28, Karl-Liebknecht-Str. 24/25, D-14476 Potsdam-Golm, Germany\\
$^{10}$INAF-Astronomical Observatory of Padova, I-36012 Asiago\\
$^{11}$Department of Physics and Astronomy, Michigan State University, East Lansing, MI 48824, USA}

\date{Accepted 2019 June 06. Received 2019 May 08; in original form 2019 February 13}

\pubyear{2018}

\begin{document}
\label{firstpage}
\pagerange{\pageref{firstpage}--\pageref{lastpage}}
\maketitle

\begin{abstract}
We report the discovery of three stars that, along with the prototype \lp, form a distinct class of chemically peculiar runaway stars that are the survivors of thermonuclear explosions. Spectroscopy of the four confirmed \lp\ stars finds ONe-dominated atmospheres enriched with remarkably similar amounts of nuclear ashes of partial O- and Si-burning. Kinematic evidence is consistent with ejection from a binary supernova progenitor; at least two stars have rest-frame velocities indicating they are unbound to the Galaxy. With masses and radii ranging between 0.20--0.28\,M$_{\sun}$  and 0.16--0.60\,R$_{\sun}$, respectively, we speculate these  inflated white dwarfs are  the partly burnt remnants of either peculiar Type\,Iax or electron-capture supernovae.  Adopting supernova rates from the literature, we estimate that $\sim$20 \lp\ stars brighter than $19$\,mag should be detectable within 2\,kpc from the Sun at the end of the {\em Gaia} mission. We suggest that as they cool, these stars will evolve in their spectroscopic appearance, and eventually become peculiar O-rich white dwarfs. Finally, we stress that the discovery of new \lp\ stars will be useful to further constrain their evolution, supplying key boundary conditions to the modelling of explosion mechanisms, supernova rates, and nucleosynthetic yields of peculiar  thermonuclear  explosions.
\end{abstract}




\section{Introduction}

\lp\ \citep[GD\,492;][]{luyten70,giclas70} is a high-velocity star unbound to the Galaxy with a unique composition: it has an O/Ne-dominated atmosphere sprinkled with the ashes of incomplete O- and Si-burning \citep{vennes17}. These authors have proposed \lp\ as a partially burnt runaway white dwarf that survived disruption by a thermonuclear supernova, specifically a peculiar group that is suggested to experience pure deflagrations,  leading to subluminous explosions  that do not completely disrupt the white dwarf accretors \citep[SN\,2002cx-like, or SNe\,Iax;][]{li03,phillips07,jordan12, foley13, kromer13, kromer15, fink14, jha17}. In \citet{raddi18a}, hereafter \citetalias{raddi18a}, we presented supportive evidence for the formation of \lp\ in a single-degenerate thermonuclear supernova by measuring a super-solar Mn abundance that is compatible with theoretical nucleosynthesis yields of near-Chandrasekhar-mass (M$_{\rm Ch}$) explosions \citep[][]{seitenzahl13a,seitenzahl17}, and it is also in agreement with the hypothesis of \lp\ as SN~Iax survivor \citep[][]{vennes17}.

In \citet{raddi18b}, hereafter \citetalias{raddi18b}, we used the precise parallax available in the Second {\em Gaia} Data Release \citep[{\em Gaia} DR2;][]{gaia18} to estimate the radius of \lp\ to be $R\simeq 0.18$\,R$_{\sun}$, i.e.\ one order of magnitude larger than canonical white dwarf radii \citep[][]{tremblay17}. We interpreted this finding with the star being currently inflated as a consequence of the supernova explosion \citep[][]{jordan12,kromer13,shen17}. The scenario of  a post-explosion expansion is compatible with the low rotational velocity of the star \citep[$v \sin{i} \leq 50$\,\kms;][\citetalias{raddi18a}]{vennes17}. Studying the kinematics of \lp, we demonstrated it is gravitationally unbound from the Milky Way \citep[see  also][]{vennes17}, having a rest frame velocity of $v_{\rm rf} \simeq 850$\,\kms\ that is about 1.5 times larger than the escape velocity at the corresponding Galactocentric radius. Considering that \lp\ has benefitted  from the Galactic rotation, we estimated it gained an ejection velocity of $\simeq 600$\,\kms\ at the moment of the supernova explosion. In the single-degenerate scenario, considering negligible contribution from asymmetric mass loss, this velocity could be achievable via the ejection from a compact binary ($1$\,hr orbital period), e.g.\ with a massive He-burning companion \citep[][]{wang09b,wang13}.

Among many exciting discoveries, {\em Gaia} DR2 has led to another possible breakthrough in the field of thermonuclear supernova research with the identification of three new hyper-runaways with $v_{\rm rf} \gtrsim 1000$\,\kms\ \citep{shen18b}. These stars form a new, relatively homogeneous spectral class of subluminous, CO-rich objects likely connected to the explosion mechanism of supernovae  Type Ia (SN\,Ia) that is known as {\em dynamically driven double-degenerate double-detonation} scenario \citep[D$^{6}$;][]{shen18a}.

\lp\ is interpreted as the formerly accreting white dwarf in a single-degenerate mass-transferring binary \citep{vennes17}, also referred to as the bound remnant. On the other hand, the D$^{6}$ stars are interpreted as the former compact donors in close, double-degenerate binaries. Also the D$^{6}$ stars are suggested to have reached their rest-frame velocities by conserving their high orbital velocities at the moment of explosion. Like \lp, the D$^{6}$ stars are currently thermally bloated; however, their expansion might be caused by the pre-explosion tidal forces and the interaction with the supernova ejecta \citep[][]{shen17}. Prior to {\em Gaia} DR2, a hyper-runaway He-rich subdwarf star, US\,708 \citep{hirsch05,heber16}, was proposed as the fastest known star that may have been the former subdwarf donor in a thermonuclear supernova event \citep{justham09,geier15}.
Following the  identification of the  D$^{6}$ stars, \citet{shen18b} have  suggested a connection with US\,708, which they proposed as representative for a possible later stage on the way back towards the canonical white dwarf cooling sequence. All together, \lp, the D$^{6}$ stars, and  US\,708 likely represent the first direct evidence of a wider class of binary progenitors for thermonuclear supernovae \citep{wang12,maoz14}.

We have performed spectroscopic follow-up of candidate high-velocity stars, which we selected by mining {\em Gaia} DR2. Our search has successfully led to the identification of two additional stars that, along with \lp, we propose to form a distinct class of partly burnt, post-supernova runaway stars (hereafter, ``\lp\ stars"). Furthermore, by mining the Sloan Digital Sky Survey (SDSS) spectroscopic database \citep{smee13,abolfathi18}, we have identified two related objects: one is very likely also an \lp\ star, while the other requires further follow-up observations.

In this manuscript, we present follow-up observations of the {\em Gaia} candidates, and the  identification of the SDSS candidates.  In Section\,\ref{sec:three}, we detail the spectral analysis\footnote{The spectra analysed here are made available at the Open Fast Star Catalogue (\url{https://faststars.space/}), which is currently maintained by Douglas Boubert and James Guillochon.} of the \lp\ stars identified with {\em Gaia}, and we re-analyse \lp\ itself including new {\em Hubble Space Telescope} ({\em HST}) near-ultraviolet (NUV) spectroscopy. We present the Galactic orbits of  the new \lp\ stars in Section\,\ref{sec:four}. In  Section\,\ref{sec:five}, we discuss  our results in a broader context, focusing on: i) the chemical composition,  compared to theoretical simulations of white dwarfs  that are  predicted to survive peculiar  thermonuclear explosions; ii) physical parameters (mass and radius); iii) evolutionary status, drawing a comparison with theoretical predictions; and iv) kinematics, birth places and binary progenitors. Finally, in Section\,\ref{sec:five-17}, we model a population of \lp\ stars to estimate  the end-of-mission detection limits of {\em Gaia} within 2\,kpc  from the Sun, deriving the distribution of astrometric parameters. In our conclusions, we  summarise  the remarkable similarities among the observed stars.

\begin{figure}
\includegraphics[width=\linewidth]{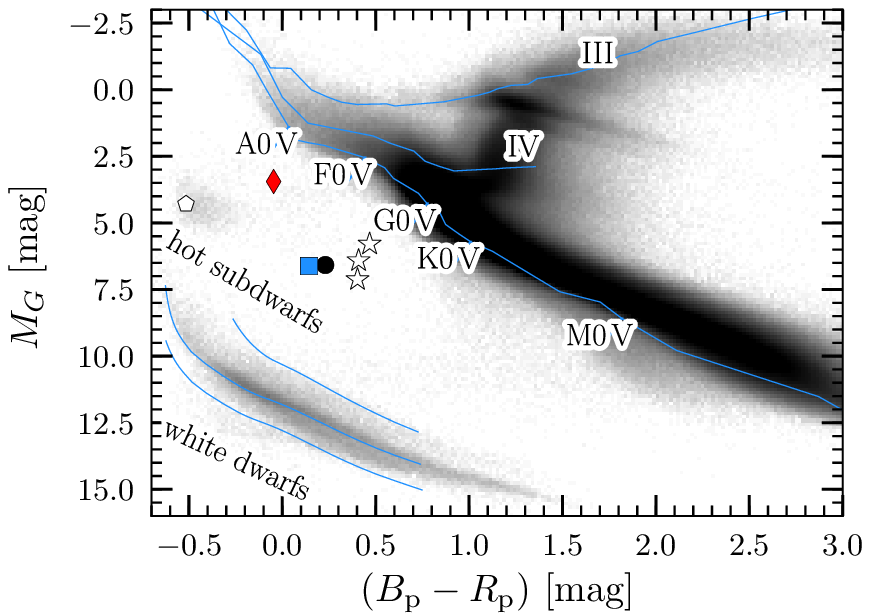}
\caption{{\em Gaia} HR diagram. \lp, \jg, and \jb\ are shown as circle, square, and diamond symbols, respectively (correspondence between colours and symbols will be maintained in the following figures). The three D$^{6}$ objects, identified by \citet{shen18b}, are marked by star-shaped symbols. US\,708, for  which we adopt a spectroscopic distance of $8.5\pm1.0$\,kpc \citep[][]{geier15}, is plotted as a pentagon symbol.
Error bars are smaller or of the order of the symbol sizes. The intrinsic colours of normal stars are interpolated from the \citet{pickles98} spectral library (luminosity class V, III, I). Cooling tracks for  0.2,\,0.6,\,1.0\,M$_{\sun}$ white dwarfs (from top to bottom) are derived by scaling a  grid of synthetic spectra in the  $T_{\rm eff} = 6000$--90,000\,K range \citep{koester10} with standard mass-radius relations for pure-hydrogen atmospheres \citep{wood95,bergeron01,fontaine01} and convolving them with the {\em Gaia} DR2 filter profiles.}
\label{f:hr}
\end{figure}

\section{Observations}
\label{sec:two}
\subsection{{\em Gaia} DR2 selection}
\label{sec:two-1}
\begin{figure}
\includegraphics[width=\linewidth]{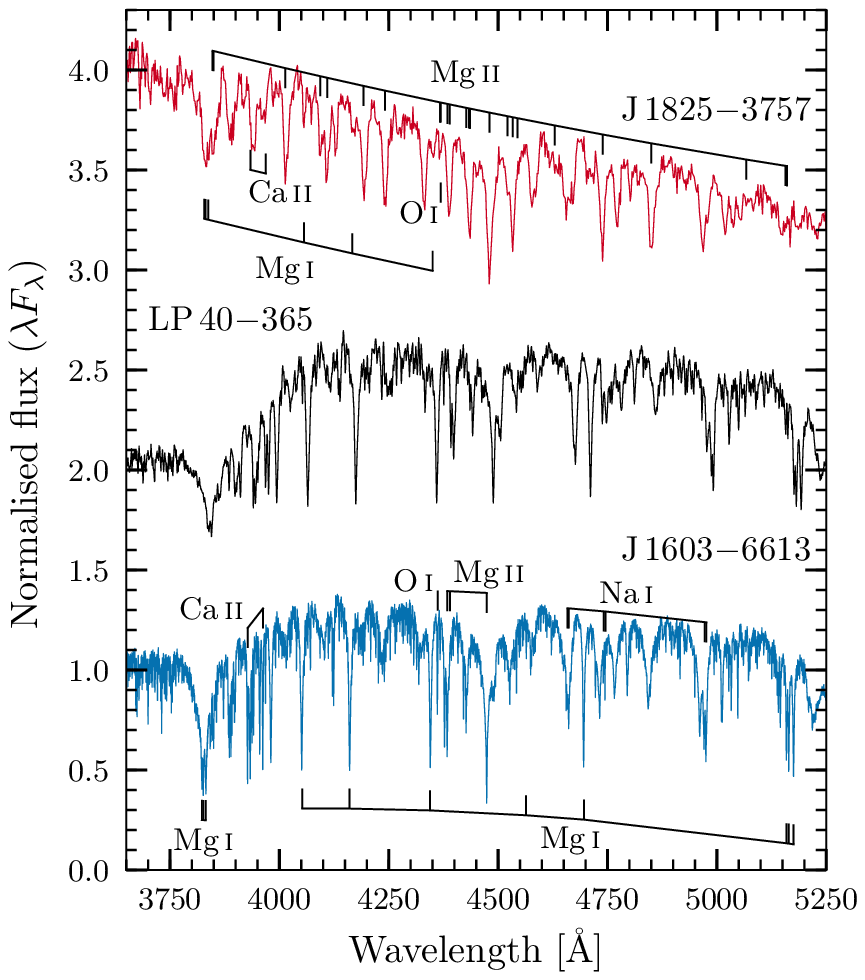}
\caption{Comparison between the new stars and \lp, with some of the strongest lines labelled. The NTT/EFOSC2 spectrum of \jb\ \citep{raddi17} and the SOAR/Goodman spectrum o \jg\ have a resolution of $\simeq 8$ and $2.8$\,\AA, respectively. The WHT/ISIS spectrum of \lp\ from \citetalias{raddi18a} has a $2$\,\AA\ resolution. All the spectra are plotted in the heliocentric frame, so spectral lines are shifted by the radial velocity  of each star.}
\label{f:spec-class}
\end{figure}

\begin{table*}
 \caption[]{Astrometric and photometric parameters
of \jg\ and \jb, including broad-band photometry from {\em GALEX} \citep[NUV;][]{morrissey07}, SkyMapper \citep[$uvgriz$;][]{wolf18}, and Vista Hemisphere Survey \citep[$J$;][]{mcmahon13}. \label{t:gaia}}
\begin{tabular}{@{}llll@{}}

\hline
Parameters & Symbols & \jg\ & \jb\ \\
\hline
Designation & & 
Gaia DR2 5822236741381879040 &
Gaia DR2 6727110900983876096\\
Right Ascension [hms] & $\alpha $ & 
16:03:04.06 &
18:25:22.15 \\
Declination [dms]& $\delta$ & 
$-$66:13:26.9 &
$-$37:57:26.1 \\
Galactic longitude [deg]& $\ell$ &
$320.63$ & 
$356.03$ \\
Galactic latitude [deg]& $b$ &
$-10.18$ & 
$-11.58$ \\
Parallax [mas]& $\varpi$ & 
$0.57  \pm 0.10$ &
$1.08  \pm 0.06$ \\
Proper motions [\masyr]& $\mu_{\alpha}\cos{\delta}$ & 
$+39.8 \pm 0.1$ &
$-37.6 \pm 0.1$ \\
& $\mu_{\delta}$ & $-7.1 \pm 0.2$ &
 $-143.0 \pm 0.1$\\
Fluxes [mag] & $G$ & $17.84$&
$13.28$\\
&$B_{\rm p}$ &  $17.89$ &
$13.21$\\
&$R_{\rm p}$  & $17.75$ &
$13.26$\\
&NUV & $20.19\pm0.22$   & $14.61\pm0.01$\\
&$u$ & $17.985\pm0.031$ & $13.251\pm0.002$\\
&$v$ & $17.829\pm0.082$ & $13.227\pm0.009$\\
&$g$ & $17.691\pm0.062$ & $13.214\pm0.006$\\
&$r$ & $17.831\pm0.030$ & $13.317\pm0.002$\\
&$i$ & $18.126\pm0.064$ & $13.609\pm0.003$\\
&$z$ & $-$              & $13.799\pm0.002$\\
&$J$ & $17.667\pm0.057$ & $13.288\pm0.002$\\
\hline
\end{tabular}
\end{table*}
We searched for high-velocity-star candidates by
selecting {\em Gaia} objects having relatively large transverse velocities ($v_{\rm t} \gtrsim 300$\,\kms) and falling within the extended colour space  below the main sequence in the {\em Gaia} colour-magnitude Hertzsprung-Russell (HR) diagram (Fig.\,\ref{f:hr}) that  is occupied 
both by \lp\ and the D$^{6}$ stars, i.e.\ delimited by {\em Gaia} colours of $B_{\rm p} - R_{\rm p} 
\lesssim 0.7$ and absolute magnitudes of $M_{G} \gtrsim 2.5$. 
Limiting the parallax precision to $\varpi/\sigma_{\varpi} > 5$, 
our criteria included \lp\ and D$6$--2, 
but excluded the two other D$^{6}$ stars. 
The spectroscopic follow-up has led to the identification of several canonical
hot subdwarfs and white dwarfs, which we will present elsewhere. 
Two stars stood out -- 
{\em Gaia}~DR2~5822236741381879040 and {\em Gaia}~DR2~6727110900983876096
(hereafter shortened to \jg\ and \jb, respectively, based on their equatorial coordinates) -- as possessing peculiar spectra, clearly resembling that of \lp\ (see Fig.\,\ref{f:spec-class}). \jg\ was previously unknown, while we previously obtained a low-resolution spectrum of \jb\ (Fig.\,\ref{f:spec-class}) on the base of which we had classified it as a likely hot subdwarf \citep{raddi17}. Follow-up observations are detailed in the next sections, and the journal of observations is listed in Table\,\ref{t:logs}.
The {\em Gaia} DR2 astrometry and photometry of the new stars are listed in Table\,\ref{t:gaia}. 

\jg\ has an apparent magnitude $G = 17.84$ that, combined with its
parallax $\varpi = 0.57 \pm 0.10$\,mas, corresponds to an absolute
magnitude $M_{G} = 6.62$, which is similar to that of \lp. This new star is 
slightly bluer than \lp, suggesting it may be hotter. The {\em Gaia} colour excess factor $(f_{BP}+f_{RP})/f_G = 1.6$ might indicate some problems with the photometry, given that the star is observed in a relatively crowded field. We note 
an interstellar extinction in the $V$-band $A_{\rm V} = 0.28$--$0.32$
\citep{sfd98,schlafly11}, i.e.\ about four times larger than the 
that along the \lp\ sightline. 

\jb\ is located closer to the Sun, 
having a parallax of
$\varpi = 1.08 \pm 0.06$\,mas. It is intrinsically bluer and about 3\,mag brighter than \lp\ and \jg, implying it is both hotter and larger than the other two stars. The total interstellar extinction is comparable
to that of \jg. 

\begin{table*}
\centering
\caption{Observing logs of the data presented and analysed in this work.\label{t:logs}}
\begin{tabular}{@{}lllll@{}}
\hline
Mode & Instrument & \lp\ &\jg\ &\jb\ \\
\hline
Spectroscopy &SOAR/Goodman & & 2018 May 6&\\
             &VLT/X-shooter & & 2018 Jul 9 & 2016 Sep 27\\
             &{\em HST}/STIS & 2018 May 21, 25& &\\
Photometry   & SOAR/Goodman & & 2018 May 10 & 2018 Jul 8\\
\hline
\end{tabular}
\end{table*}

\subsection{\texorpdfstring{\jg}{J1603-6613}}
\label{sec:two-2}
\subsubsection{Low-resolution spectroscopy}
We first observed \jg\ on 2018 May 6 with the Goodman Spectrograph \citep{clemens04} mounted on the 4.1\,m Southern Astrophysical Research (SOAR) telescope at Cerro Pach\'on in Chile. We obtained 8 $\times$ 300\,s consecutive spectra using a 930 line/mm grating with wavelength coverage between $3600-5200$\AA. We used a $1\farcs01$ slit and bracketed our exposures with a 60\,s Fe arc lamp, since no skylines exist in this spectral region for wavelength calibration. With a dispersion of 0.84 \AA/pixel our resolution is roughly 2.8\,\AA. 

The spectra were optimally extracted \citep{horne86} using standard {\sc starlink} routines \citep[][]{currie14}, after applying bias and quartz-lamp flat corrections using the software {\sc pamela}. We used the software package {\sc molly} \citep{marsh89} to wavelength calibrate the spectra, apply a heliocentric correction, and perform a final weighted average of the one-dimensional (1D) spectra, and used the spectrophotometric standard EG\,274 for flux calibration. We did not detect any radial velocity shift between the exposures, within the precision of the wavelength calibration, and thus coadded all spectra to obtain a spectrum with a signal-to-noise ratio of ${\rm S/N} \simeq 20$ per pixel, shown in Fig.\ref{f:spec-class}. 
Noting the striking similarity with \lp, we used the strongest lines to
estimate the radial velocity with respect to the Sun via cross correlation
with our synthetic model of \lp, obtaining $v_{\rm rad} = -480 \pm 
10$\,\kms\ that confirmed \jg\ as a high-velocity star. 
\subsubsection{Intermediate-resolution spectroscopy}
We followed up \jg\ again on 2018 July 9, with the multi-wavelength
intermediate-resolution spectrograph X-shooter \citep{vernet11}, mounted at the Cassegrain focus of the ESO Very Large Telescope (VLT) 
UT2 in Cerro Paran\'al (Chile). We binned the detectors ($1 \times 2$ in
the spatial and dispersion directions, respectively) and we
used narrow slits of $1\farcs0$,
$0\farcs9$, and $0\farcs9$, in the UVB, VIS, and NIR arms, respectively, 
taking six exposures in each arm 
and achieving total exposure times of 7320, 7500, and 7800\,s, in 
order to reduce the impact of cosmic rays and to confirm the absence of radial-velocity variability. The mean seeing was of $0\farcs6$.

The data were reduced using the standard ESO pipeline Reflex \citep{reflex}, and the \texttt{molecfit} software \citep{molecfit1,molecfit2} was used to remove the 
telluric absorption of Earth's atmosphere. Again, we did not observe 
any significant radial-velocity variation over short timescales, so we coadded the individual exposures to obtain an average spectrum, 
confirming the same blue-shifted radial velocity. The coadded spectrum
has a ${\rm S/N} \sim 100$ in the blue/visual spectral range,
with resolving power $R = 5400$, $8900$, and $5600$, in the UVB, VIS,
and NIR arms, respectively.  

\subsubsection{Time-series photometry}
We performed time-series photometry of \jg\ on the night of 2018 May 10 using the imaging mode of the Goodman spectrograph on SOAR. 
We used an exposure time of $17$\,s and had a readout time of $2.0$\,s, and observed for $2.7$\,hr through a \textit{S8612} broad-bandpass, red-cutoff filter (3300--6200\,\AA). 
Seeing across the duration of the observation was fairly stable, averaging $1\farcs$--$1\farcs5$.

Images were debiased and flat-fielded using an internal dome lamp. We extracted a light curve using standard circular aperture photometry, with a 4.5-px aperture surrounded by an 8-px annulus. The target light curve was normalised using a bright, nearby comparison star with no close companions, and we subtracted out a second-order polynomial to account for airmass changes. A Lomb-Scargle periodogram showed no significant peaks above $0.2$\,per\,cent amplitude for periods between roughly $40$--$4000$\,s. This lack of variability suggests that \jg, like \lp, is consistent with not-having an unseen close companion. 

\subsection{\texorpdfstring{\jb}{J1825-3757}}
\label{sec:two-3}

\subsubsection{Intermediate-resolution spectroscopy}
In \citet{raddi17} we published a low-resolution spectrum (8\,\AA) of
\jb\ taken at the New Technology Telescope with the ESO Faint Object
Spectrograph and Camera v.2 \citep[NTT/EFOSC2;][]{buzzoni84} 
at the La Silla observatory (shown
in Fig.\ref{f:spec-class}). Although we classified this star as
a likely hot subdwarf,  its spectrum shows numerous unusually strong 
\ion{Mg}{i-ii} lines and other ionised metals. The 
absence of H and He lines resembles the spectral 
appearance of \lp\ and \jg.

We followed up \jb\ on 2016 September 27 with VLT/X-shooter, taking single  exposures of 521, 550, and
600\,s in the blue, visual, and near-infrared arms, respectively. The instrument  setup and  data  reduction steps were the same as those  adopted for \jg. The mean seeing during the observations of \jb\ was  $0\farcs7$. The achieved resolving power were 4300, 7400, and 5400, in the UVB, VIS, and NIR, respectively. The extracted spectrum has a ${\rm S/N}> 150$ in the UVB/VIS arms and $>100$ in the NIR arm. 
\subsubsection{Time-series photometry}
We also observed \jb\ through a Bessel-V filter on the night of
2018 July 08 with the imaging mode  of the Goodman spectrograph on SOAR. We collected $1140\times5$\,s exposures, and each exposure had roughly 2.0\,s of dead time from readout. Seeing was steady from $1\farcs6$--$1\farcs7$. As with \jg\ we bias- and flat-field corrected the photometry and extracted the light curve using a fixed 5.5\,px circular aperture, and subtracted a second-order polynomial to account for airmass changes throughout the 2.2\,hr run. We did not see any coherent variability at any period from $15$-$3500$\,s, to a limit of at least 0.1\,per\,cent amplitude, consistent with it currently being an isolated star.
\subsection{\texorpdfstring{\lp}{LP 40-365}}
\label{sec:two-4}
\subsubsection{HST spectroscopy}

We observed \lp\ with {\em HST} using the Space Telescope Imaging
Spectrograph \citep[STIS;][]{woodgate98} on 2018 May 21 and 25, 
during Cycle
25, for three and two orbits, respectively. The total exposure time 
amounted to 15\,253\,s. We used the G230L grating and the NUV
Multianode Microchannel Array (MAMA) detector with a 
$52\arcsec \times 2\arcsec$ aperture, which delivers a
resolving power $R \simeq 1000$ across the 1570--3180\,\AA\ wavelength 
range. We applied the wavelength and flux calibration via standard {\sc iraf} {\sc calstis} tasks \citep{katsanis98}.
We measured an average S/N$ \simeq 35$ from the reduced spectrum.

\subsection{SDSS spectra}
\label{sec:two-5}

We also considered the possibility that similar stars may reside within the 4.9 million spectra of the SDSS database \citep[][]{abolfathi18}, which could have been previously misclassified due to their unusual spectral features.

We adapted the template method of \citet{hollands17}, which essentially fits all SDSS spectra against a library of spectral templates. We created a grid of synthetic \lp\ spectra, i.e.\ with \Teff\ ranging from 5000 to 7000\,K in steps of 500\,K, and then up to 20\,000\,K in steps of 1000\,K. In all cases the $\log g$ was fixed at 5\,dex, and the abundances fixed to the average of the already identified stars (as justified by the results presented  in Section~\ref{sec:three}). For each model, additional spectral templates were offset by fixed radial-velocity shifts ranging from $-1500$ to $+1500$\,\kms, in steps of 50\,\kms, resulting in a total of 1098 templates. Finally, we convolved each template with a Gaussian to match the SDSS spectral resolution of 2.7\,\AA. To filter other easily identifiable spectral types, we complemented the grid of \lp\ templates with 1843 white dwarf templates (spectral classes DA, DB, DZ), and 1045 stellar templates of spectral types commonly observed by SDSS (see description in \citealt{hollands17}). 
For each SDSS spectrum, we interpolated all templates onto the SDSS wavelength scale, flux-scaled them to the data, and then evaluated their $\chi^2$, adopting the template with the lowest $\chi^2$ as the best match. The scaling factor was the only free-parameter, whose optimal value was determined analytically  \citep[see eq.\,1 in][]{hollands17}. The closest-matching template, the corresponding reduced $\chi^2$ ($\chisqr$), and the mean spectral signal-to-noise (S/N) between 4500--6000\,\AA\ were recorded for each SDSS spectrum.

\begin{figure}
\includegraphics[width=\columnwidth]{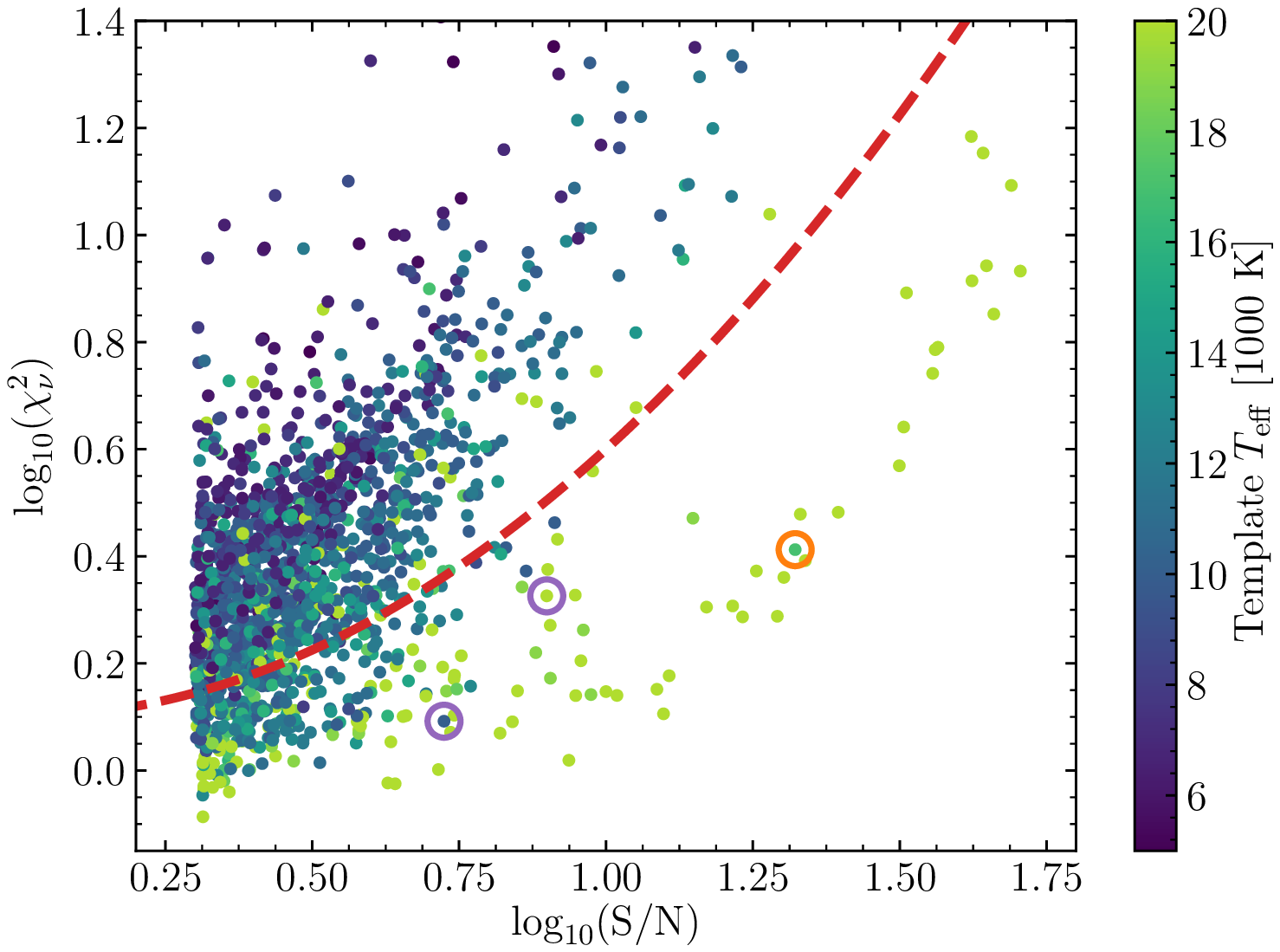}
\caption{\chisqr\ vs. spectral S/N for SDSS spectra that match best to an \lp\ template. The red dashed curve indicates a cut, below which the corresponding spectra were visually inspected. Two new 
candidates are circled in purple, whereas \dox\ is circled  in orange.}
\label{fig:chisn}
\end{figure}

Out of the 3.1 million SDSS spectra with a mean $\mbox{S/N} > 2$, 1544 were best matched by a \lp\ model. Even so, a closest match did not guarantee a good fit. In Figure~\ref{fig:chisn}, we show the \chisqr\ of the best-matching templates vs. the SDSS spectrum mean S/N, with the template-\Teff\ indicated by the point-colours. Towards high S/N, the distribution in \chisqr\ becomes bimodal. Points in the upper-branch correspond to spectra where all templates provided poor matches, but with one of the \lp\ templates as the least bad. The lower-branch corresponds to more convincing matches to the templates. Notably, most of the points in the lower branch correspond to high-\Teff\ templates, as their metal-lines become weaker, and thus the number of false-positives increases. The red-dashed line indicates our cut in the \chisqr-S/N plane. The spectra corresponding to the 349 objects below the cut were visually inspected and compared with their best fitting templates.

\begin{figure*}
\includegraphics[width=\linewidth]{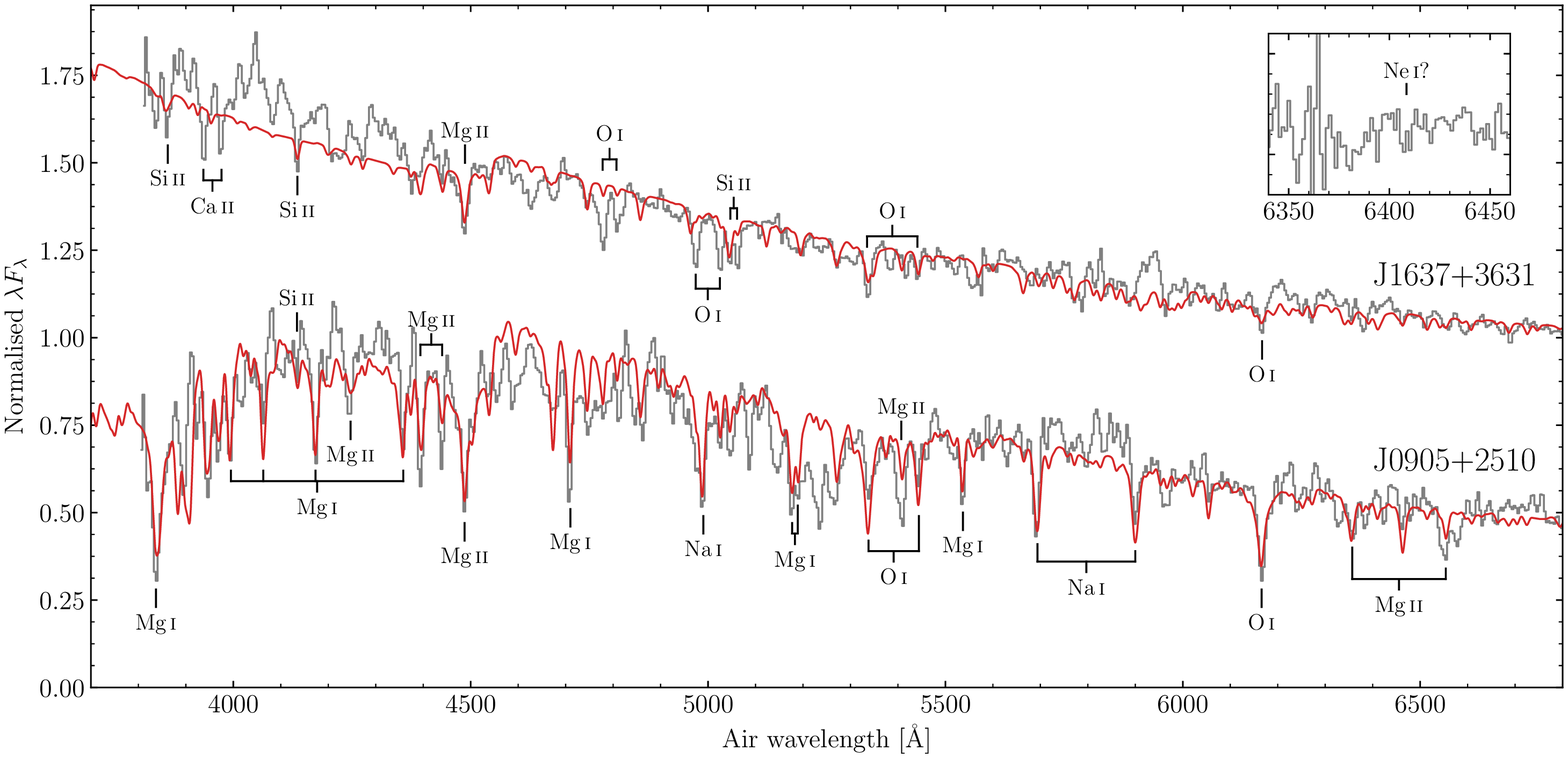}
\caption{Normalised SDSS spectra of \jc\ (bottom) and \jd\ (top, offset by $+0.75$). Their best matching ``\lp\ star'' templates are shown in red. For clarity, the spectra and templates have been convolved by Gaussians of $\mbox{FWHM}=8$\,\AA\ to better show the strongest transitions (labelled). For \jd, an inset shows a tentative \Ion{Ne}{i} detection (at the nominal SDSS resolution). Both spectra are shown in the heliocentric frame.}
\label{fig:sdss_spectra}
\end{figure*}

Following this visual inspection, three spectra showed convincing resemblance to their templates. One of these (Fig.\,\ref{fig:chisn}, circled in orange) is the O-dominated atmosphere white dwarf (DOx), \sdss{124043.00}{+}{671034.6} \citep[hereafter \dox;][]{kepler16}, where matches to Mg, O, and Si lines resulted  in  a favourable \chisqr. The other two objects (Figure~\ref{fig:chisn}, circled in purple) are \sdss{090535.55}{+}{251011.3} and \sdss{163712.21}{+}{363155.9}, (hereafter \jc\ and \jd\ respectively). The spectra of the two SDSS candidates, with their best-matching templates, are shown in Figure~\ref{fig:sdss_spectra}. Their astrometry and photometry are provided in Table~\ref{tab:sdss}.

\begin{table*}
 \caption[]{\label{tab:sdss} Astrometry, photometry, and spectroscopic parameters for the objects identified from SDSS spectra, \jc\ and \jd. Reddening values were determined from the Pan-STARRS dust-maps \citep{chambers16,green18}. Photometric $\Teff$ were calculated from fitting our ``\lp\ star'' models to the GALEX NUV and SDSS photometry including the quoted reddening. Results from our template matching are given at the bottom.}
\begin{tabular}{@{}llll@{}}

\hline
Parameter [units] &  Symbol & \jc\ & \jd\ \\
\hline
Gaia Designation   &&  Gaia DR2 688380457507044864 & Gaia DR2 1327920737357113088 \\
Right Ascension [hms]  &$\alpha$  & 09:05:35.55  & 16:37:12.21 \\
Declination   [dms]     &$\delta$  &  $+$25:10:11.3      & $+$36:31:55.9 \\
Galactic longitude [deg & $\ell$ &  201.65 &  58.98 \\
Galactic latitude [deg] &$b$& $+39.71$ &  $+41.82$ \\
Parallax    [mas]       & $\varpi$&  $-$   &  $0.39\pm 0.60$ \\
Proper motions$^{1}$ [\masyr] 
& $\mu_\alpha\cos\delta$ &  $3.1 \pm 3.4$  & $-18.5\pm1.2$ \\
&$\mu_\delta$ &  $24.7 \pm 3.0$    & $+64.2\pm1.2$\\
Fluxes      [mag]  &$G$&  $19.62$ & $20.33$\\
                   &$B_{\rm p}$& $19.69$ & $20.19$\\
                   &$R_{\rm p}$& $19.36$ & $20.21$\\
                   &NUV& $22.23\pm0.28$ & $20.44\pm0.17$ \\
                   &$u$& $20.26\pm0.05$ & $20.01\pm0.05$\\
                   &$g$& $19.67\pm0.01$ & $20.09\pm0.02$\\
                   &$r$& $19.73\pm0.02$ & $20.40\pm0.04$\\
                   &$i$& $19.96\pm0.03$ & $20.62\pm0.06$\\
                   &$z$& $20.03\pm0.12$ & $20.86\pm0.28$\\
Interstellar reddening [mag]           &$E(B-V)$& $0.0\pm0.02$ 
& $0.02\pm0.02$ \\
Photometric temperature [K]   &\Teff\ & $10\,050\pm140$ & $13\,500\pm600$ \\
\hline
Template parameters     &\Teff\ [K]& $10\,000$      & $20\,000$ \\
       &RV [\kms]& $+300$      & $+300$ \\
         &\chisqr\ & 1.24       & 2.12 \\
Spectrum S/N             && 5.30                         & 7.93 \\
Plate-MJD-fiber          && 2086-53401-0500              & 2185-53532-0360 \\
\hline
\multicolumn{4}{l}{Notes: $^{1}$The proper motions of \jc\ are taken from the Pan-STARRS--SDSS cross-match \citep{tian17}.}
 \end{tabular}
   \end{table*}
\subsubsection{\texorpdfstring{\jc}{J0905+2510}}
With a mean S/N of 5.3, a cursory inspection of the SDSS spectrum of \jc\ simply appears as a noisy blackbody. However, degrading the resolution to 8\,\AA\ (Figure~\ref{fig:sdss_spectra}, bottom) reveals lines of \Ion{O}{i}, \Ion{Na}{i}, \Ion{Mg}{i/ii}, and \Ion{Si}{ii}. The close agreement between the strong lines in the data and best-matching template indicates that this star also belongs to the class of ``\lp\ stars.'' We also note some discrepancy near 5230\,\AA, corresponding to a feature, currently not accounted  for by the models, that is also observed in \lp\ as well as in \jg\ and \lp\, which \citet{vennes17} attributed to a resonance in the photoionisation cross-section of Mg.

\jc\ has only a two-parameter astrometric solution (position) within the {\em Gaia} DR2 data. With $G = 19.62$, it is probable that a parallax will be available in a future {\em Gaia} data release. Assuming \jc\ has an absolute $G$ magnitude similar to \lp\ and \jg\ (6.5 mag), the true distance is likely to be about $4$\,kpc from the Sun. Given the proper motion of $(\mu_\alpha \cos\delta, \mu_\delta) = (3.1\pm 3.4,\ 24.7\pm 3.0)$\,\masyr \citep{tian17}, the implied transverse velocity is about 500\,\kms, which is broadly consistent with the 300\,\kms\ radial velocity.

To more accurately measure the \Teff, we fitted our \lp\ model grid to {\em GALEX}~NUV and SDSS photometry (interpolating amongst models for estimating intermediate \Teff\ to our grid-steps). We accounted for interstellar reddening, $E(B-V) = 0.04\pm0.02$\,mag, \citep[as estimated via the Pan-STARRS 3D dust-maps at a heliocentric distance of at 4\,kp;][]{green18}, 
obtaining $9840\pm170$\,K. A full abundance analysis and comparison with the other members of this class necessitates higher-quality follow-up spectra.

\subsubsection{\texorpdfstring{\jd}{J1637+3631}}
This star was first reported as a white dwarf with a C-rich, He-dominated  atmosphere (DBQ spectral type) in the SDSS DR7 white dwarf catalogue of \citet{kleinman13}. However, we neither see evidence for He nor C features. Beyond the general Rayleigh-Jeans slope, the main spectral features matched by the template (Figure~\ref{fig:sdss_spectra}, top) are the 4129\,\AA\ \Ion{Si}{ii} and 4481\,\AA\ \Ion{Mg}{ii} lines (rest wavelengths), although redshifted by about 300\,\kms. Transitions from \Ion{O}{i} and \Ion{Ca}{ii} appear stronger than in the template, indicating somewhat different abundances than assumed for our grid of models. The 6402\,\AA\ \Ion{Ne}{i} line is also tentatively detected in the unsmoothed SDSS spectrum (Figure~\ref{fig:sdss_spectra}, inset). 

For \jd, {\em Gaia} DR2 provides a full five-parameter astrometric solution, although rather poorly constrained. With a parallax of $0.39\pm0.60$\,mas, we place a 3\,$\sigma$ lower limit of about 500\,pc on the distance. The implied absolute magnitude is $M_G = 11.8$. Applying a smooth exponentially decreasing prior that accounts for the stellar density distribution  in the Milky Way,  \citet{bailer-jones18} have estimated a distance of  $d = 1550^{+1115}_{-630}$\,pc, which implies an intrinsic magnitude  of $M_G = 9.3\pm 1.2$. Both  these distance estimates place \jd\ within the white dwarf sequence of Figure~\ref{f:hr}. 

Of course \jd\ could be much further away and still remain consistent with the {\em Gaia} parallax. If we assume \jd\ is a \lp\ star, and conservatively, like \lp\ itself, has $M_G=6.5$, the implied distance would be about 6\,kpc. Given the 67\,\masyr proper-motion (Table~\ref{tab:sdss}), the transverse velocity would be $\sim$2000\,\kms\ -- several times larger than theoretical predictions for SNe~Iax survivors \citep{shen18b}.

Hence, a more likely explanation is that \jd\ could be the  second representative of the DOx white dwarf class, after \dox\ \citep{kepler16}. This hypothesis can be tested with medium-resolution follow-up spectroscopy.
One difference we note between these two stars is the intensity of the \Ion{Ca}{ii} H+K lines: they are plainly visible in the spectrum of \jd, but only an upper-limit of Ca was determined for \dox. Our photometric \Teff, determined from fitting {\em GALEX} and SDSS photometry ($13\,500\pm600$\,K, Table~\ref{tab:sdss}), is cooler than the $21\,590\pm620$\,K measured\ by \citet{kepler16}  for \dox, which could explain the differences in \Ion{Ca}{ii} strength.

\section{Spectral analysis}
\label{sec:three}

\subsection{\texorpdfstring{\jg}{J1603-6613}}
\label{sec:three-1}

\begin{figure*}
\includegraphics[width=\linewidth]{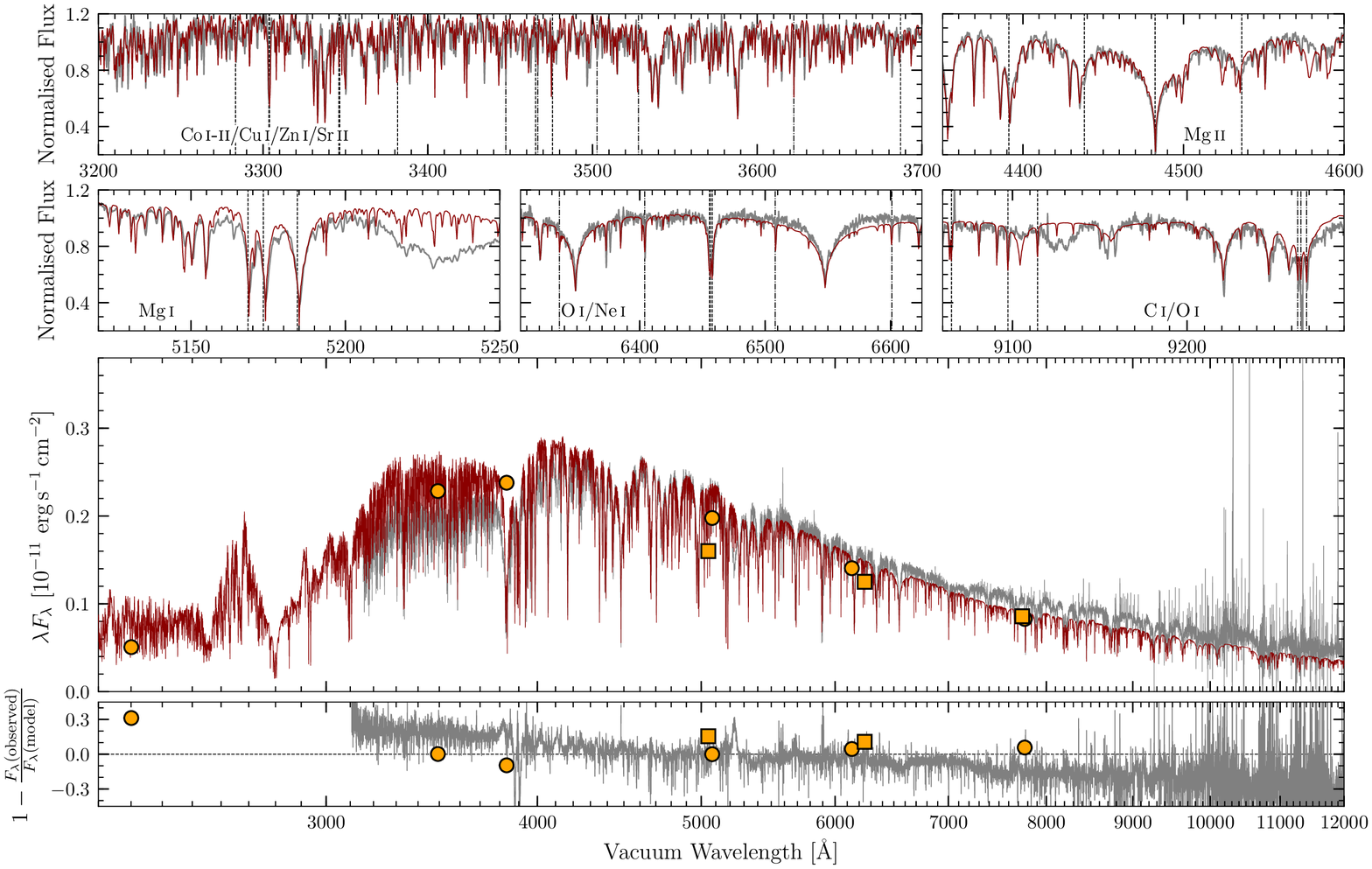}
\caption{Comparison between the X-shooter spectrum of \jg\ (grey) and the best-fit synthetic spectrum (dark red). {\em Top panels:} The $u$-band region displaying absorption from Co, Cu, Zn, and Sr; spectral ranges containing representative examples of Mg\,{\sc i-ii}, O\,{\sc i}, Ne\,{\sc i}, and C\,{\sc i} lines. {\em Middle panel:} De-reddened spectral energy distribution. The observed {\em GALEX} NUV and Sky-Mapper $ugri$ data are over-plotted as circles, while the {\em Gaia} $B_{\rm P}$, $G$, and $R_{P}$ magnitudes are represented by squares (see  Table\,\ref{t:gaia}). {\em Bottom panels:} Residuals between observed and synthetic spectra, and observed and synthetic photometry. The best-fit  model was obtained by minimising the residuals between observed and synthetic photometry, hence the  slope  observed for the observed vs. synthetic spectrum comparison.
\label{f:j1603_sed}}
\end{figure*}
\begin{figure}
\includegraphics[width=\linewidth]{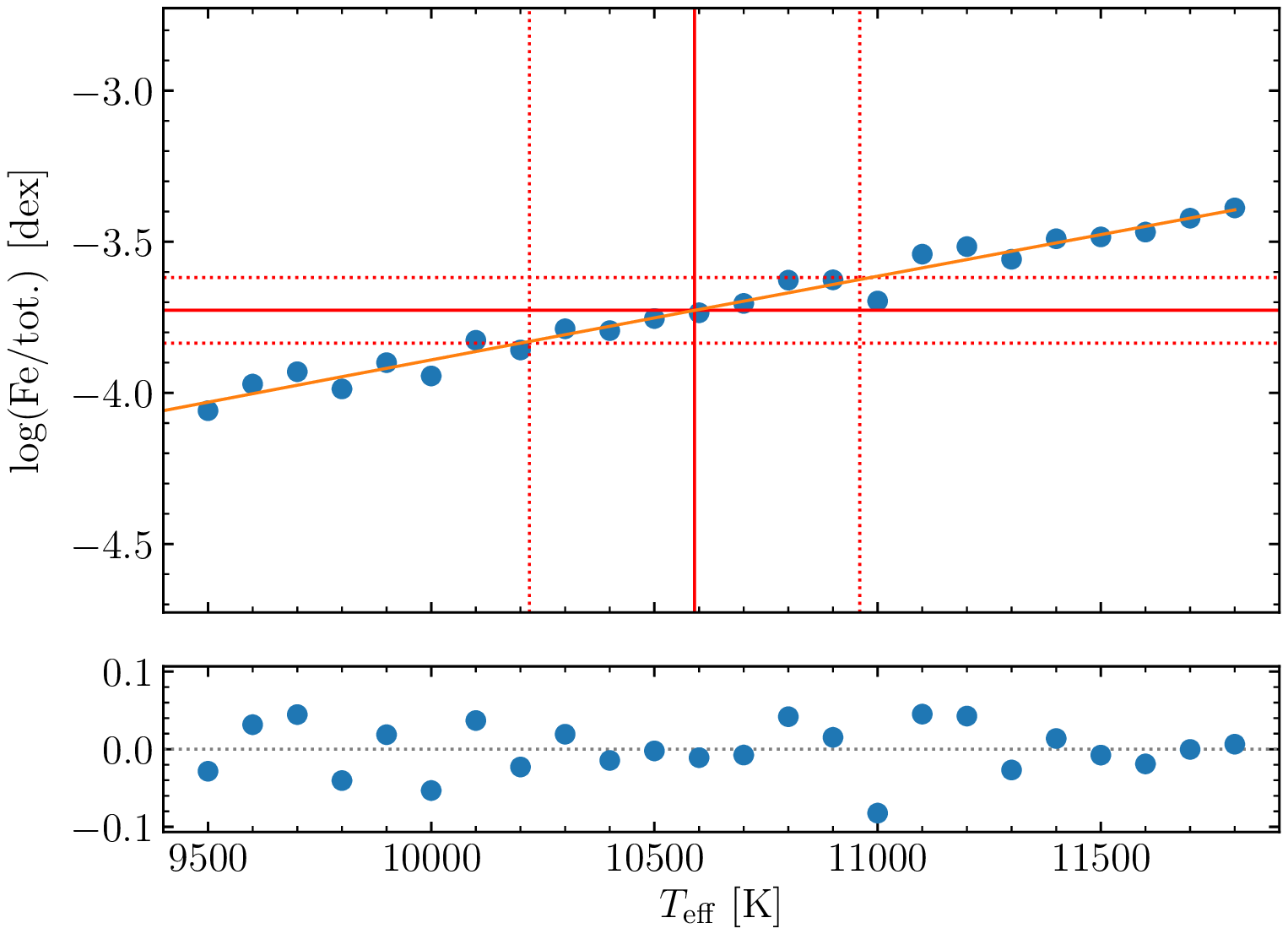}
\caption{Estimation of the Fe abundance uncertainty for \jg. The top panel shows the fit (orange) to the optimal abundance at each \Teff\ point. The bottom panel shows the residuals with standard deviation of $0.04$\,dex. The variation of 0.10\,dex over the $370$\,K \Teff\ uncertainty leads to $-3.73\pm0.11$\,dex overall. \label{f:1603_Fe}}
\end{figure}

Fig.\,\ref{f:spec-class} shows that \jg\ has a spectrum remarkably similar to \lp, thus providing a natural starting point for the spectral analysis. Owing to the high S/N and  resolution of the X-shooter spectrum, we searched  for lines from other elements in addition to those previously identified in \lp. We identified new lines from \Ion{Sc}{ii}, \Ion{Co}{i--ii}, \Ion{Zn}{i}, \Ion{Cu}{i}, and \Ion{Sr}{ii}. \Ion{C}{i} lines were also found to be present at optical wavelengths. As with \lp, no evidence for H or He were seen, but with more stringent limits from the much higher S/N ratio and  $\simeq 1000$~K hotter \Teff.  

We precisely measured the blue-shift of \jg\ by fitting Lorentzian profiles to some of the strong, isolated O~{\sc i} and Mg\,{\sc i} lines in the VIS arm of  the  X-shooter spectrum, obtaining $v_{\rm rad} = -485 \pm 4$\,\kms, which confirms the estimate from the identification spectrum.

As previously noted in \citetalias{raddi18a} for \lp, 
a hotter $T_{\rm eff}$ clashes with the non-detection of \ion{He}{i} absorption at 5877\,\AA. Given the similarity between the two stars, we reconsidered  the feasibility of a ONe-dominated atmosphere, as found  by \citet{vennes17} for \lp.
Our model spectra were computed following the methods of \citet{koester10}, as already done  in \citetalias{raddi18a}. We have made use of the most recent atomic
data\footnote{The most relevant updates for the current project 
include new oscillator strengths and 
broadening constants from the atomic line
databases NIST (National Institute of Standards and Technology;
Kramida et al. 2019), VALD \citep[Vienna Atomic Line Data;][]{ryabchikova15,kupka00}, and Robert Kurucz's line  lists  for \ion{O}{i}  and \ion{Mg}{ii} (\url{http://kurucz.harvard.edu/}). Also new is the inclusion of the
absorption of the negative ions of Ne, O, Na
\citep{liu99a,liu99b,robinson67,john75a,john75b,john75c,john96}.}. 
One important result is that also ONe-atmospheres develop relatively deep convection zones ($\sim  10^{-6}$\,M$_{\rm star}$), which extend down to the bottom of the model atmosphere like for He-dominated atmospheres in the same temperature range. Although we could not estimate the size  of the  mixing zone below the atmosphere, we have estimated the diffusion timescales of the detected elements to be, at  least, of the order of 10\,Myr (this result will  be  relevant  for  the discussion in Section\,\ref{sec:five-12}).

We initially approached the spectral analysis by fitting  the  X-shooter spectra alone. Given that O, Ne, and Mg, are the dominant elements and main contributors to the continuum opacity, their abundances  are necessarily
related to each other and coupled  with $T_{\rm eff}$.
Hence, due to the inherent inaccuracy of flux calibration and the ill-constrained effect of interstellar reddening, 
the resulting atmospheric fit was strongly degenerate, in a way that adjustments in \Teff\ could be compensated for by changes in abundances, allowing for reasonable agreement with the data over a wide \Teff-range. 
We therefore tried a different approach.

Instead of allowing \Teff\ as a free parameter, we performed multiple least squares fits to the normalised spectra, but at fixed \Teff\ in steps of 100\,K in the range 9500\,K to 11\,800\,K. The optimal $\log{g}$ and abundances varied across the grid steps. We then used this model grid to fit SkyMapper and {\em GALEX} photometry (Table\,\ref{t:gaia})
with \Teff, the interstellar reddening and stellar radius as free parameters, and the \emph{Gaia} parallax and its error used as a Gaussian prior. While Gaia photometry were also available, their quoted uncertainties do not include systematic errors, which are expected to be significantly larger than this. Thus, we chose to exclude these. The best-fit was achieved by calculating synthetic photometry in each bandpass and minimising the \chisqr. This resulted in a \Teff\ of $10590\pm370$\,K  and a radius of $0.16\pm0.03$\,R$_{\sun}$. While the full line-of-sight reddening is about $E(B-V)=0.10$, our fit indicated a 99~per~cent upper limit of $0.09$ with a mean value of $0.03$ in the posterior distribution, indicating the full line-of-sight extinction to be too high. 

To obtain estimates for $\log g$, abundances, and their errors, we fitted the \Teff\ dependence of the  element, $Z$, with a 2nd-order polynomial. The standard deviation of the scatter around the polynomial was taken as $\sigma_{Z_0}$, and the gradient determined at $\Teff=10590$\,K. Finally, the uncertainty on $Z$ was calculated as

\begin{equation}
    \sigma_Z^2 = \sigma_{Z_0}^2 + \left(\frac{\partial Z}{\partial \Teff}\right)^2\,\sigma_{\Teff}^2,
\end{equation}
where $\sigma_{\Teff} = 370$\,K. We show this graphically for  Fe as an example (Fig.\,\ref{f:1603_Fe}). We compare the  observed  spectrum and  our best-fit model in Fig.\,\ref{f:j1603_sed}. The  atmospheric  parameters and the element abundances are listed in Table\,\ref{tab:fits_new}.

\begin{figure*}
\includegraphics[width=\linewidth]{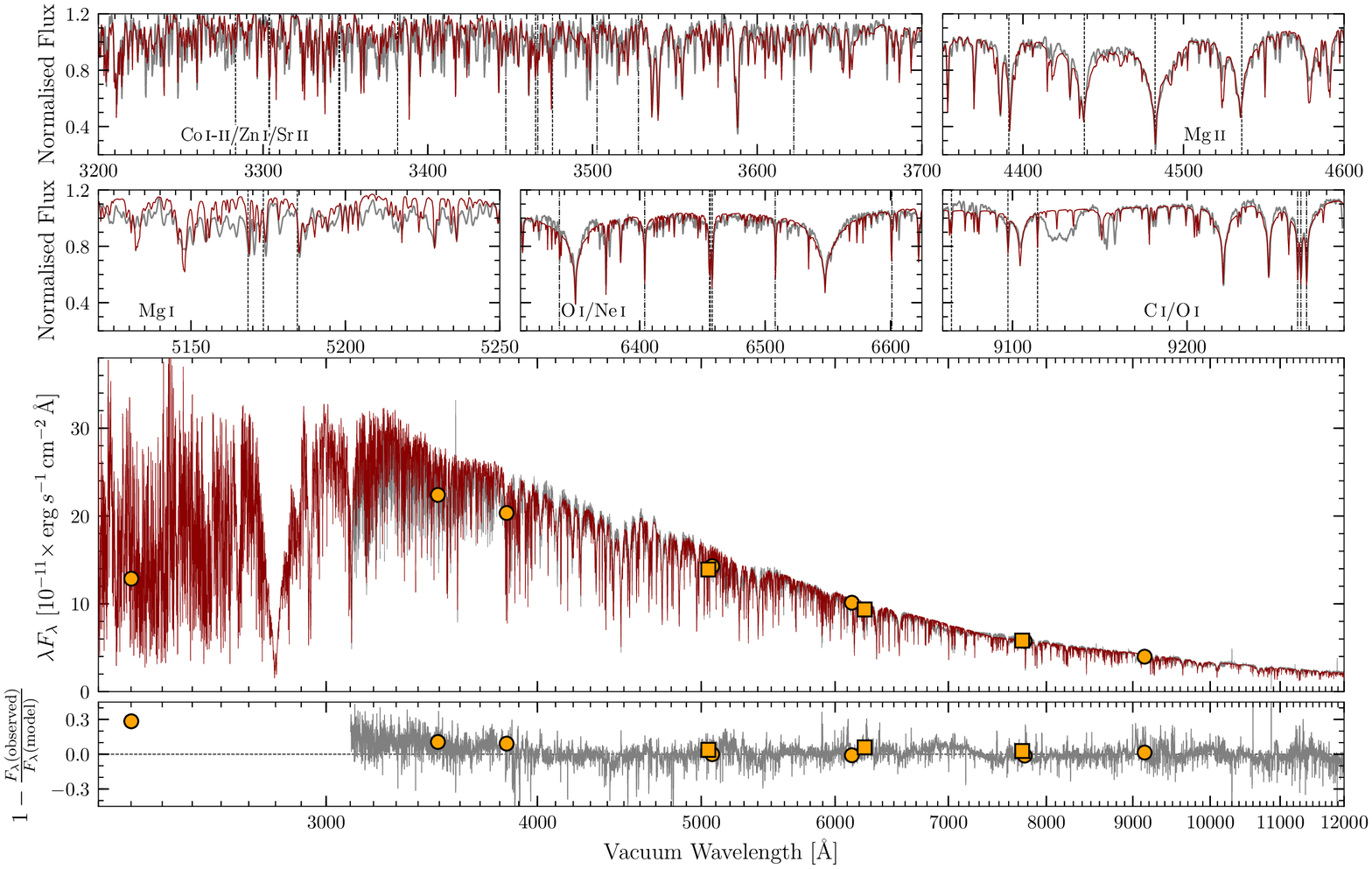}
\caption{The X-shooter spectrum, photometry, and best-fit model for \jb. Colours and symbols match those in Fig.\,\ref{f:j1603_sed}.
\label{f:j1825_sed}}
\end{figure*}

\subsection{\texorpdfstring{\jb}{J1825-3757}}
\label{sec:three-2}

We confirmed the radial velocity of \jb, obtaining $v_{\rm rad} = -47\pm3$\kms\ from Lorentzian fits to isolated O\,{\sc i}, Mg\,{\sc ii}, and Ca\,{\sc ii} lines of  the  X-shooter spectrum.

Once again we searched the spectrum to identify all elements present in the stellar atmosphere.
As inferred from the bluer $B_{\rm p}-R_{\rm p}$ colour, \jb\ is hotter than both \lp\ and \jg, and so
the ionisation balance for each element is expected to be different.
Apart from Cu (the higher \Teff\ does not permit detection of \Ion{Cu}{i}, and no
strong optical lines of \Ion{Cu}{ii} are expected), all elements found
at \lp\ and \jg\ were re-identified.
We  note  that several other elements populate higher
ionisation states than found for the other two stars: \Ion{Al}{iii} was identified from doublets
around 3610\,\AA\ and 5710\,\AA, \Ion{Si}{iii} from a doublet around 4560\,\AA,
\Ion{O}{ii} at 4350\,\AA\ and 4650\,\AA, and \Ion{S}{ii} from a multitude of lines.

For  the  spectral analysis, we took a similar approach to that
described for \jg. We  built a model grid between $\Teff\
11\,800$--$14\,000$\,K, in 100\,K steps, with the surface gravity and abundances as free parameters.
For the photometric fit we made use of the SkyMapper and 2MASS data (Table\,\ref{t:gaia}). We excluded \textit{GALEX} photometry from our fit, because the NUV spectral region is expected to be more heavily line blanketed than found for the optical,  and  we cannot rule out unaccounted opacity sources, e.g. from elements presently unidentified.
Using the {\em Gaia} parallax (Table\,\ref{t:gaia}) as a prior,
we found $\Teff =12830\pm450$\,K, $E(B-V) = 0.080\pm0.017$,
and $R = 0.60\pm0.03\,R_\odot$. 

The values and uncertainties on $\log g$ and abundances were calculated in the same manner as described for \jg, 
and are given in Table~\ref{tab:fits_new}.
Note that, because of the higher \Teff, the  upper-limits for hydrogen and helium are far more constraining than for \lp\ and \jg. This strongly affirms the ONe-dominated nature of these stars.
We compare the observed spectrum and our best-fit model in Fig.\,\ref{f:j1825_sed}.

A major difference in the results of \jb\ vs. \lp\ and \jg\ is the surface
gravity, which is more than one order of magnitude lower. This is not too surprising given
the much larger radius we found for the photometric fit, implying the star is more inflated compared to the other two. This is discussed further in  Section\,\ref{sec:five-13}.

As a final comment, we examine the properties of the
updated atomic data that have a more tangible effect
in modelling the spectrum of \jb. 
While our line list already included many thousands of
transitions from the NIST and VALD3 databases, we noticed that the profiles  of some of the  weakest \ion{O}{i} and \ion{Mg}{ii} lines  were poorly reproduced  by our models. In particular, three moderately strong lines at 4014, 4094,
and 4110\,\AA\ (rest-frame air wavelengths) 
were absent in our model. From their intensities and large line widths we concluded these are most likely
higher members of the \ion{Mg}{ii} series transitioning from the closely spaced
4d/f levels (93\,311/93\,800\,cm$^{-1}$,  respectively), as 
confirmed from wavelengths tabulated by \citet{moore59}.
We found the three missing lines (and some higher members of the series that are less obviously present)
in Robert Kurucz's line archive, 
which includes not only
wavelengths and oscillator strengths, but also Stark, quadratic Stark, and van der Waals
broadening constants. 
This update in our models shows a close agreement
with the observed spectrum. However, we note that for the highest series members, the
strengths are systematically under-predicted, likely owing to the proximity of the upper-levels to
the ionisation limit, adversely affecting their estimated occupation numbers. Nevertheless, the Mg  abundance was already precisely constrained by other well-known, stronger  Mg II lines. The updated \Ion{Mg}{ii} lines are  generally weaker  in the spectra  of \jg\ and  \lp, because of their lower \Teff. 
Hence, their inclusion in the models leads to a minor improvement. 

Although most  of the strong lines  are well determined  in the  literature, our concluding remark is that, because of their extreme photospheric composition, \lp\ stars necessitate complete and accurate atomic data even for lines usually considered astrophysically irrelevant.

\subsection{\texorpdfstring{\lp}{LP 40-365}}
\label{sec:three-3}
\begin{figure}
\includegraphics[width=\linewidth]{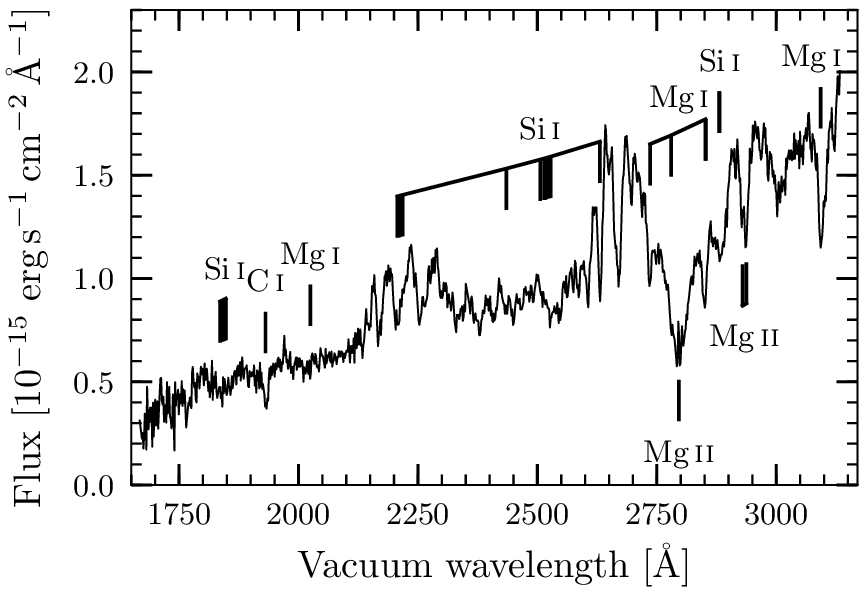}
\caption{{\em HST}/STIS spectrum of \lp. The red-shifted spectrum has been corrected into the laboratory rest-frame. We note the C\,{\sc i} detection and the strongest Mg and Si lines, which cause most of the opacity in the spectrum. \label{f:hst}}
\end{figure}
\begin{table}
\centering
 \caption[]{\label{tab:fits_new} Composition our best-fitting synthetic spectra of \lp\ (Section\,\ref{sec:three-1}), \jg\ (Section\,\ref{sec:three-2}), and \jb\ (Section\,\ref{sec:three-3}). Abundances are expressed as the number fraction of the total, $\log{(\rm{Z}/\rm{Tot}})$, where the contribution from is not included.}
\begin{tabular}{@{}lccc@{}}
 \hline
Parameters & \lp & \jg\ & \jb\ \\
\hline
\Teff\,[K]      & $9800 \pm 350 $ & $10\,590\pm370$ & $12\,830\pm450$ \\
$\log g$\,[cgs] & $5.5\pm0.3    $ & $  5.34\pm0.20$ & $  4.21\pm0.18$ \\
$v_{\rm rad}$ [\kms]   & $499 \pm 6$ & $-485 \pm 5$ & $-47 \pm 3$ \\
\hline
H               &  $< -4.30     $ & $ < -5.8      $ & $ < -6.5      $ \\
He              &  $<-0.60      $ & $ < -1.8      $ & $ < -3.5      $ \\
C               &  $-2.80\pm0.25$ & $ -3.00\pm0.22$ & $ -3.08\pm0.08$ \\
O               &  $-0.46\pm0.23$ & $ -0.45\pm0.08$ & $ -0.44\pm0.03$ \\
Ne              &  $-0.27\pm0.15$ & $ -0.22\pm0.05$ & $ -0.21\pm0.02$ \\
Na              &  $-2.10\pm0.26$ & $ -2.02\pm0.11$ & $ -2.28\pm0.07$ \\
Mg              &  $-1.20\pm0.32$ & $ -1.60\pm0.07$ & $ -1.67\pm0.04$ \\
Al              &  $-2.40\pm0.25$ & $ -2.84\pm0.32$ & $ -3.05\pm0.08$ \\
Si              &  $-2.80\pm0.25$ & $ -3.40\pm0.17$ & $ -2.96\pm0.05$ \\
P               &  $<-4.30      $ & $ < -5.5      $ & $ < -6.4      $ \\
S               &  $-3.20\pm0.25$ & $ -4.05\pm0.24$ & $ -3.96\pm0.14$ \\
Ca              &  $-5.20\pm0.25$ & $ -5.52\pm0.11$ & $ -4.87\pm0.12$ \\
Sc              &  $<-5.40      $ & $ -6.22\pm0.26$ & $ -7.63\pm0.35$ \\
Ti              &  $-4.70\pm0.34$ & $ -6.15\pm0.19$ & $ -6.51\pm0.11$ \\
V               &  $<-4.40      $ & $ < -7.0      $ & $ < -9.2      $ \\
Cr              &  $-4.40\pm0.35$ & $ -5.23\pm0.22$ & $ -5.45\pm0.09$ \\
Mn              &  $-5.00\pm0.35$ & $ -5.16\pm0.18$ & $ -5.14\pm0.05$ \\
Fe              &  $-3.00\pm0.25$ & $ -3.73\pm0.11$ & $ -3.50\pm0.03$ \\
Co              &  $<-3.90      $ & $ -4.65\pm0.19$ & $ -4.14\pm0.08$ \\
Ni              &  $-3.80\pm0.35$ & $ -4.04\pm0.11$ & $ -4.15\pm0.06$ \\
Cu              &  $<-4.30      $ & $ -5.50\pm0.30$ & $ -           $ \\
Zn              &  $<-5.30      $ & $ -5.20\pm0.20$ & $ -5.50\pm0.30$ \\
Sr              &  $<-7.10      $ & $ -8.14\pm0.29$ & $ -8.40\pm0.25$ \\
\hline
\end{tabular}
\end{table}
\begin{figure*}
\includegraphics[width=\linewidth]{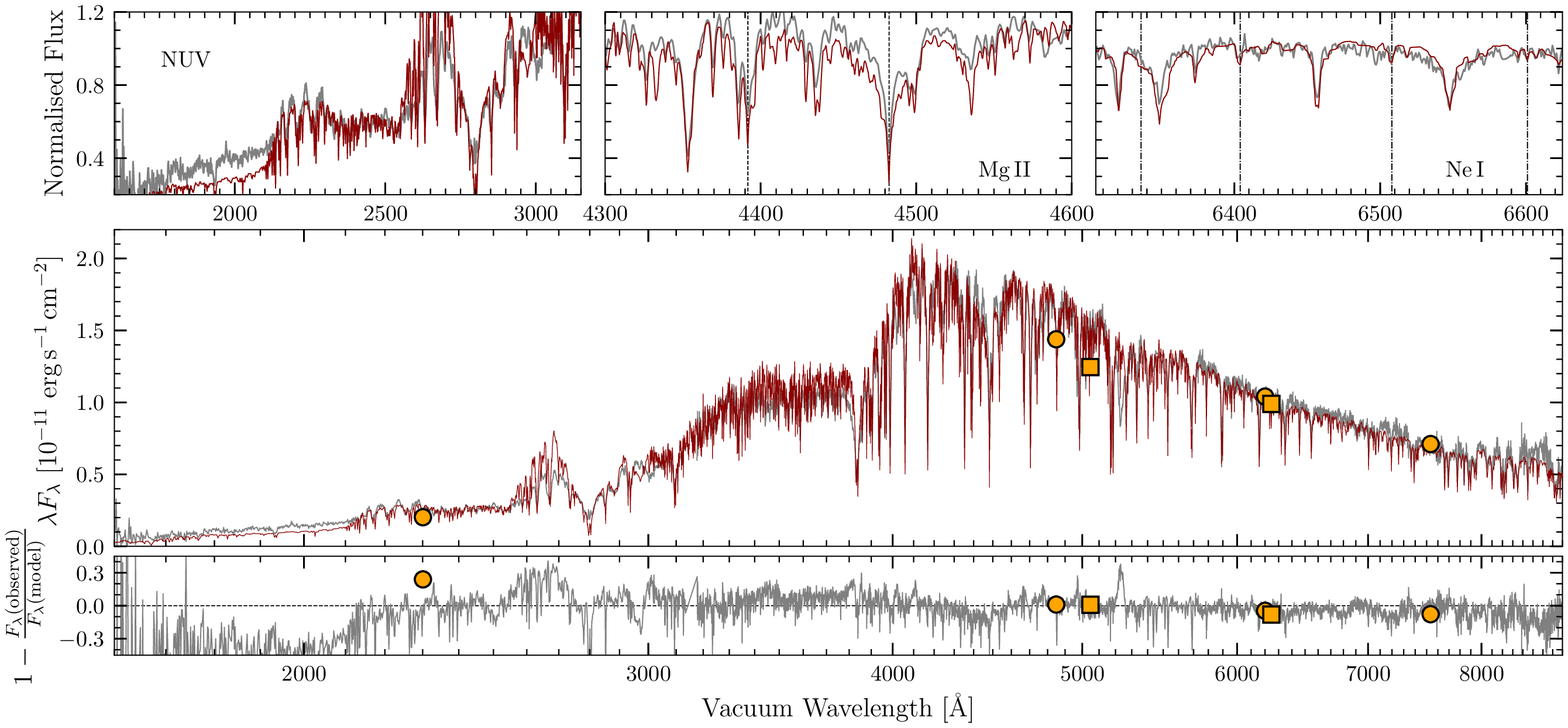}
\caption{Comparison between the combined  NUV/optical spectrum of \lp\ (grey) and the best-fit synthetic spectrum.
Colours and symbols match those in Fig.\,\ref{f:j1603_sed}
and \ref{f:j1825_sed}.
{\em Top panels:} Spectral ranges containing the NUV spectrum (left), the optical regions with strong \ion{Mg}{ii} transitions (centre) and \ion{Ne}{i} lines (right), both marked by dashed vertical lines. {\em Middle panel:} De-reddened spectral energy distribution with the {\em GALEX} NUV, Pan-STARRS $gri$, and  {\em Gaia} photometry over-plotted. {\em Bottom panels:} Residuals between observed and synthetic spectra, and observed and synthetic photometry. 
\label{f:lp40_sed}}
\end{figure*}

We re-analysed the prototype star, combining our  new
$HST$/STIS spectrum (Fig.\,\ref{f:hst}) 
with the WHT/ISIS and Asiago 1.82-m Copernico/AFOSC spectra from \citetalias{raddi18a}.
This combined, flux-calibrated spectrum has an average spectral resolution of $\simeq 2$--$4$\,\AA\ in the optical and NUV ranges, respectively. The 5350--5700\,\AA\ region, which is only covered by the
Copernico/AFOSC data, has a 13\,\AA\ resolution. 
We scaled the optical spectrum to the Pan-STARRS magnitude \citep[$g = 15.635$;][]{flewelling16}. Although the $HST$/STIS spectrum seamless matches the optical fluxes, we note it is 4\,$\sigma$ brighter than the {\em GALEX}~NUV magnitude. The optical spectrum, instead, appears slightly bluer than the published Pan-STARRS and {\em Gaia}~DR2 photometry, corresponding to a small colour term in the range of $B - V = -0.02$, which is comparable to the photometric uncertainties.
In joining the spectra together, we transposed the optical region onto a vacuum wavelength scale to match the $HST$/STIS data \citep[using the relation given in][]{morton00}. We corrected  the combined  spectrum for radial velocity  shift of \lp\ ($v_{\rm rad} = +500$\,\kms), which is  confirmed  by the redshift  of  the  \ion{C}{i} line at 1930\,\AA\ and other strong  \ion{Mg}{i} lines.
Finally, we corrected the combined spectrum for 
the total interstellar reddening of
$E(B-V) = 0.026$ \citep{sfd98}, 
corresponding to $A_{g} \simeq 0.09$ and $A_{\rm NUV} \simeq 0.22$ magnitudes of extinction in the
Pan-STARRS~$g$ and {\em GALEX}~NUV bands, 
respectively, as estimated via the
\citet{fitzpatrick99} standard $R_{V} = 3.1$ 
extinction law. The combined dereddened spectrum is displayed in Fig.\,\ref{f:lp40_sed}.

From an initial appraisal of the combined spectrum, 
the NUV data indicate a hotter temperature with 
respect to the optical spectrum, for which we 
inferred $T_{\rm eff} = 8900$\,K using a He-dominated atmosphere in \citetalias{raddi18a}. In the previous sections, we have shown the high-quality X-shooter spectra  of \jg\ and \jb\ are well matched by synthetic spectra with ONe-dominated atmospheres. Especially from the analysis of \jb, we have  excluded the  presence of He  with a very stringent limit. Hence, these results 
favour an ONe-dominated atmosphere also for \lp,
as found by \citet{vennes17}.

As we did  for \jg\ and \jb, we  built a grid of synthetic models in the $T_{\rm eff} = 8900$--$10\,200$\,K range, with O, Ne, and Mg as  dominating species. 
In view of the complex nature of the problem, we approached the analysis of \lp\  by identifying a set of key spectral features to compare against, in order to evaluate the goodness of fit and  to evaluate the  impact  of systematic uncertainties. These are: i) the level of NUV flux; ii) the spectral lines of Ne, O, and Mg; iii) the continuum slope in the optical range; iv) the \ion{Mg}{i} jump near 3800\,\AA; and v) the ionisation equilibrium of \ion{Mg}{i}/\ion{Mg}{ii}. Because we did not find a solution that satisfies all the above criteria, we have relied more on the flux-calibrated NUV spectrum, obtaining a best-fit synthetic spectrum with $T_{\rm eff} = 9800$\,K and $\log{g} = 5.5$ that is shown in Fig.\,\ref{f:lp40_sed}. This model approximately reproduces the NUV flux between 2200--3000\,\AA, the \ion{O}{i}, \ion{Ne}{i}, and \ion{Mg}{i} lines, and the \ion{Mg}{i} jump. The main defect of the synthetic spectrum is that it does not accurately reproduce the ionisation equilibrium of \ion{Mg}{i}/\ion{Mg}{ii}, over-predicting the intensity of \ion{Mg}{ii} lines as shown in the top-centre panel of Fig.\,\ref{f:lp40_sed}. In addition, we also note that the residuals between the observed and synthetic spectra, displayed in the lower panel of Fig.\,\ref{f:lp40_sed}, suggest that the optical continuum of the synthetic spectrum is slightly too steep, whereas the flux below 2200\,\AA\ is $\sim 30$~per~cent too low. As  noted in \citetalias{raddi18a}, models with increasing  lower  $T_{\rm eff}$ incur in another problem, because they have stronger \ion{Ne}{i} lines that overpredict the observed ones. 

Based on the strong coupling between O, Ne, and Mg, we estimated a $1\sigma$ $T_{\rm eff}$ uncertainty of $350$\,K, beyond which the absorption caused by these elements becomes too strong or too weak. The surface gravity cannot be constrained more precisely than $\log{g} = 5.5 \pm 0.3$\,dex, which is compatible with a $\simeq0.16$\,R$_{\odot}$ radius, as derived from the {\em Gaia} parallax and Pan-STARRS $g$ magnitude, and by assuming a plausible mass between $0.15$--$0.60$\,M$_{\sun}$ (a more detailed discussion, with reference to \citetalias{raddi18a}, is given in Section\,\ref{sec:five-12}).

The $HST$/STIS observations have enabled a better overview of the composition of \lp, clarifying the incompatibility between a He-dominated atmosphere proposed in \citetalias{raddi18a} and an ONe-dominated atmosphere, as determined by \citet{vennes17}. This result agrees with the analysis of \jg\ and \jb. The $HST$/STIS data favour a best-fit model with $T_{\rm eff} = 9800$\,K that is now more compatible with that determined by \citet{vennes17}. We also note that the dominant elements of our best-fit model are  in the ratio of O:Ne:Mg\,$\sim$\,$34$:$54$:$6$, compared to O:Ne:Mg\,$\sim$\,$50$:$40$:$2$  determined  by \citet{vennes17}

The NUV spectrum has allowed us to detect the \ion{C}{i}~1930 \,\AA\ line, enabling us to constrain the abundance of  this element. Thus, we have determined abundances for 14 elements (C, O, Ne, Na, Mg, Al, Si, S, Ca, Ti, Cr, Mn, Fe, Ni). The abundances of detected trace elements were measured by iteratively altering them, one by one, and keeping the abundances of the other elements fixed. The mean values and errors are established from the means and standard deviations derived by fitting the strongest lines of each element. We have confirmed a remarkable similarity between element abundances for the three studied  \lp\ stars (more details are given in Section\,\ref{sec:five-11}). 
Finally, accounting for the coupling between \Teff\ and element  abundances, we estimated the systematic uncertainties affecting our abundance determinations. On one hand, $\log{{\rm (O/Ne)}}$ slightly anti-correlates  with \Teff, increasing by 0.05 dex for a 350~K decrease in temperature. On the other hand, $\log{({\rm Mg/Ne})}$  strongly correlates with \Teff, decreasing by $0.4$\,dex for a $-350$~K variation. All the other trace elements follow a similar trend as  Mg. This result  implies their mutual number abundance ratios are almost unchanged with respect to Fe, although the overall bulk atmospheric composition is affected (see Section\,\ref{sec:five-12}).

We have also estimated upper limits on H, He, P, V, Sc, Co, Cu, Zn, and Sr. We note  that Sc, Co, Zn, and Sr are detected in \jg\ and \jb, and Cu is detected in \jg. The abundances with $1\sigma$ uncertainties or upper limits are listed in Table\,\ref{tab:fits_new}.

To conclude, as discussed in the previous sections and  in \citetalias{raddi18a}, it is likely that important physics may be still missing from our models.  In addition to the already mentioned atomic line data,  we note the lack of absorption coefficients for the bound-free and free-free contribution of Mg$^{-}$, Al$^{-}$, and Si$^{-}$ ions could also explain unidentified spectral features.

\section{Kinematic analysis}
\label{sec:four}
\begin{figure}
\includegraphics[width=\linewidth]{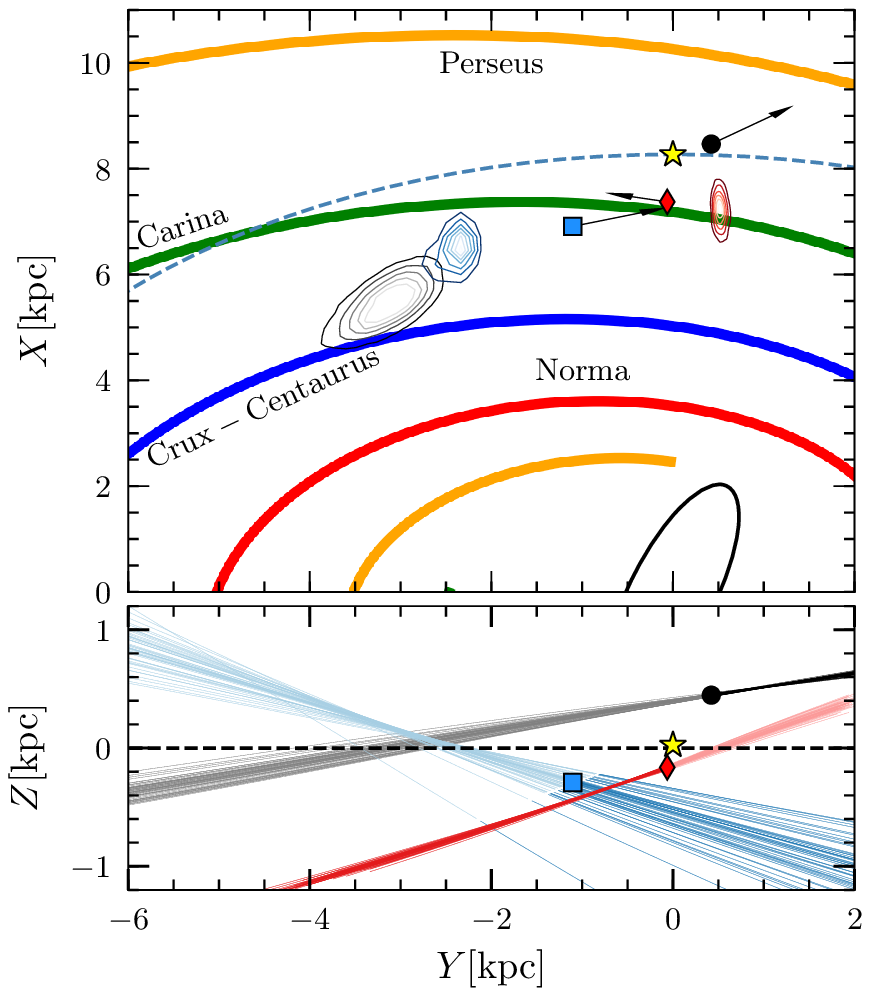}
\caption{Sampled Galactic trajectories of \lp\, \jg, and \jb. In both panels, the Sun, \lp, \jg, and \jb, are represented as a yellow star, black circle, blue square, and red diamond, respectively. None of  the  three stars crosses  the Galactic plane close  to its  centre, excluding any ejection mechanism  that requires the interaction with the super-massive  black hole, Sgr~A$^{*}$. {\em Top panel:} Top-down view of the Galactic disc. The locations of spiral arms and the Galactic bar are shown as coloured and black curves \citep[][]{vallee08}. The Solar circle is traced by a dashed curve. The arrows show the mean motions of the four stars. The grey, blue, and red contours represent the Galactic disc crossing locations of \lp, \jg, and \jb, respectively. {\em Bottom panel:} Representation of simulated trajectories in the $Y$-$Z$ plane. The simulated trajectories are plotted as grey, blue, and red lines, for \lp, \jg, and \jb, respectively, using lighter/darker tones for past/future positions. \label{f:tr}}
\end{figure}

Following the procedure described in \citetalias{raddi18b} for \lp, we used {\em Gaia} DR2 astrometry and $v_{\rm rad}$ to trace,  in a  probabilistic fashion,  the motion of \jg\ and \jb\ within the Milky Way potential. Using the {\sc python} package for galactic dynamics, {\sc galpy} \citep{bovy15}, we modelled the Milky Way's potential with the standard module \verb|MWPotential2014|, which includes a power-law density profile with exponential cut-off for the bulge, a Miyamoto-Nagai disc, and a dark matter halo \citep[see details in][]{bovy13,bovy15}.

We imposed a Galactic disc rotation of $V_{\rm c} = 239 \pm 9$\,\kms, with the Sun located at $R_{0} = 8.27 \pm 0.29$\,kpc \citep{schonrich12}, accounting for a one-to-one correlation between $R_{0}$ and $V_{\rm c}$, and a peculiar motion as determined by \citet{schondrich10}. We sampled 10\,000 boundary conditions with a Monte Carlo method, by assuming Gaussian distributions for the Galactic parameters, the radial velocities, and the astrometric parameters of the two stars. For the latter we took into account the full covariance matrix as delivered in {\em Gaia} DR2. The orbital parameters of \jb\ and \jg\ are summarised in Table\,\ref{t:kin}, where we also list those of \lp\ for comparison.

Given the lack  of or the low precision of 
{\em  Gaia} parallaxes for the two SDSS stars, 
\jc\  and \jd,
we did not compute their detailed  orbits.
Nevertheless,
we note that the radial and transverse velocities  of  \jc, which we suggest to be another \lp\ star (see  Section\,\ref{sec:two-5} for  details), deliver a rest-frame  velocity of $v_{\rm rf} \sim 730$\,\kms, making it likely unbound from the Milky Way.

\subsection{The unbound trajectory of \texorpdfstring{\jg}{J1603-6613}}
\label{sec:four-1}

Although the parallax of \jg\ is less precise than that of \lp, we can confirm it as unbound from the Milky Way with a rest frame velocity of $v_{\rm rf} = 805^{+49}_{-32}$\,\kms, exceeding the  escape  velocity of $v_{\rm esc} = 550$\,\kms\ estimated with \verb|MWPotential2014| at the  corresponding Galactocentric radius of  $R_{\rm G} = 7$\,kpc \citep[cf with $\approx520$--$530$\,\kms\ in the  Solar  neighbourhood;][]{piffl14,williams17}. We show the Galactic trajectories of the three \lp\ stars with precise {\em Gaia} parallaxes in Fig.\,\ref{f:tr}. \jg\ is escaping from the Milky Way in the southern Galactic hemisphere, while \lp\ is departing from the disc at positive latitudes. \jg\ has not yet reached its closest approach to the Sun; its spectrum is blue-shifted. Both stars move along the direction of the Galactic rotation, diverging from the tangential vectors by $\sim20$--26\,deg on average. They also have grazing trajectories with respect to the plane ($\simeq 5$ and $10$\,deg for \lp\ and \jg, respectively), which imply a significant contribution of the Galactic rotation to the rest-frame velocity. 

\begin{table*}
 \caption[]{Orbital parameters of \lp, \jg, and \jb \label{t:kin}.}
\begin{tabular}{@{}llrrr@{}}
\hline
Parameters & Symbols & \lp\ & \jg\ & \jb\ \\
\hline
Distance [kpc]& $d$ & $0.63\pm0.01$ &
$1.77^{+0.40}_{-0.27}$&
$0.93 \pm 0.05$\\
Galactocentric radius [kpc]& $R_{\rm G}$ & $8.48^{+0.30}_{-0.28}$&
$7.00^{+0.35}_{-0.38} $&
$7.37^{+0.27}_{-0.31} $\\
Elevation [kpc]& $Z$ & $0.45\pm0.01$ &
$-0.29^{+0.05}_{-0.07}$ &
$-0.16 \pm 0.01$\\
Rest frame velocity [\kms]& $v_{\rm rf}$ & $852\pm11$&
$ 804^{+49}_{-32}$ &
$408^{+39}_{-34}$\\
Eccentricity& $e$ & 
1 &
1 &
$0.69 \pm 0.12$\\
Vertical component of angular moment [kpc\,\kms]& $J_{Z}$ & $7220\pm250$ &
$5650 \pm 310 $&
$-2800 \pm 280$\\
Maximum elevation [kpc]& $Z_{\rm max}$ & $-$&
$-$&
$7.2^{+5.8}_{-2.6}$\\
Pericentre [kpc]& $R_{\rm peri}$ & $6.15\pm0.25$ &
$6.87\pm0.31$&
$7.0 \pm 0.30$\\
Apocentre [kpc]& $R_{\rm apo}$ & $-$&
$-$&
$39^{+31}_{-14}$\\
Azimutal period [Myr]& $T{\rm p}$ & $-$&
$-$&
$730^{+270}_{-660}$\\
Escape probability [\%]& & 
100 &
100 &
<10\\
Flight time from plane crossing [Myr]& $\tau_{\rm fl}\,(Z = 0) $ & 
$5.3\pm0.5$ &
$1.66\pm0.13$ &
$1.48 \pm 0.02$ \\
Galactocentric radius at plane crossing [kpc]& $R_{\rm G}\,(Z = 0)$ & 
$6.3\pm0.3$& $6.9 \pm 0.30$ & $7.2\pm0.3$ \\
Distance at plane crossing [kpc] & $d\,(Z = 0)$ & $4.2^{+0.5}_{-0.2}$& $3.0^{+0.3}_{-0.2}$& $1.28^{+0.08}_{-0.07}$ \\

\hline
\end{tabular}
\end{table*}

We note that the two unbound stars may have crossed the Galactic plane at mutually close locations, as displayed in Fig.\,\ref{f:tr}, although with different flight times from the plane, $\tau_{\rm fl}\,(Z=0)$. We computed the mutual separations between the past trajectories of \lp\ and \jg\ as a function of time, finding that the  minimum separation of 1\,kpc might have occurred 1\,Myr  ago. The orbit integration confirms that their similar spectral appearance is a characteristic of the formation channel and  its not related  to a specific event that produced both stars at the same  time.

\subsection{\texorpdfstring{\jb}{J1825-3757} is on a Milky Way bound orbit}
\label{sec:four-2}
Combining the {\em Gaia} parallax and proper motions with the radial velocity of \jb, we estimate a rest frame velocity of $v_{\rm rf} = 408^{+39}_{-34}$\,\kms, which is $\simeq 4 \sigma$ slower than $v_{\rm esc} \approx 560$\,\kms\ estimated at $R_{G} = 7.4$\,kpc. This implies that \jb\ is currently gravitationally bound to the Milky Way. For this star, we followed the orbit evolution up to 10\,Gyr from now, in order to estimate the average eccentricity ($e$), the pericentre and apocentre, the azimuthal period, and the escape probability, which we estimate as the fraction of orbits reaching $R_{G} = 100$\,kpc. These and other relevant kinematic parameters are listed in Table\,\ref{t:kin}. 

We note that \jb\ is relatively close to the pericentre of the predicted orbits, and it is  moving  towards low Galactic coordinates, i.e. negative  $Z$. It goes around the Milky Way once every $\simeq 730$\,Myr, moving on an eccentric ($e = 0.69$), halo-like orbit against Galactic rotation. If \jb\ had previously crossed the Galactic plane, it would have occurred $\simeq 1.5$\,Myr ago. Its relatively large but negative vertical component of the angular moment, $J_{Z} = -2800$\,kpc\,\kms, is consistent with its orbit being rather flat ($Z_{\rm max} \simeq 7$\,kpc). We also note that the  simulated orbits reach quite  large large Galactic radii (apocentre at $R_{G} = 39^{+31}_{-14}$\,kpc). A small fraction of our simulated orbits ($< 10$~per~cent) imply a possible escape of \jb\ from the Milky Way in the next 10\,Gyr.

\section{Discussion}
\label{sec:five}

The evidence presented so far demonstrates that \lp, \jg, \jb, and  likely \jc, have unique -- but mutually similar -- spectral characteristics and peculiar kinematics, which make them clearly distinct from other classes of stars. Our new observations strongly support the interpretation that the new stars, like  \lp, are the partly burnt white dwarf accretors that survived disruption from a thermonuclear supernova in a single-degenerate scenario \citep[possibly a SNe\,Iax, as initially advanced by][and further reiterated in our \citetalias{raddi18a,raddi18b}]{vennes17}. 
In the following  sections, we will discuss the properties that identify these stars as members of what appears  to  be a class of ``\lp\ stars.'' We will focus on the three stars with precise {\em Gaia} parallaxes for which we performed a detailed spectral analysis, but our conclusions can be extended to \jc, which just awaits {\em Gaia} to measure its  parallax and proper-motions as well as a higher-quality spectrum for a detailed abundance analysis.
\subsection{Physical properties}
\label{sec:five-1}
\subsubsection{Atmospheric composition}
 \label{sec:five-11}
\begin{figure}
\includegraphics[width=\linewidth]{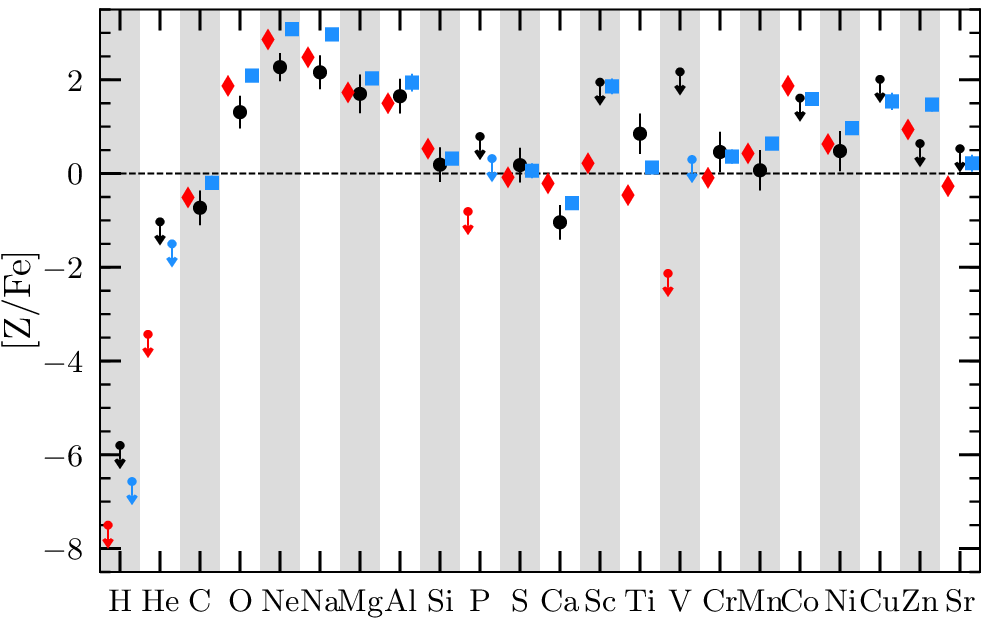}
\caption{The elemental number abundances of \lp, \jg, and \jb, detailed in Table\,\ref{tab:fits_new}, are displayed relatively to Fe and normalised to the Solar composition \citep[, dashed line about 0;][]{asplund09}. Black circles, blue squares, and red diamonds represent \lp, \jg, and \jb, respectively. When error bars are not visible, their size is comparable to or smaller than the symbol. Upper limits are shown as downward pointing arrows. \label{f:abd}}
\end{figure}
The comparison among the atmospheric composition of the three \lp\ stars  is displayed in Fig.\,\ref{f:abd}, where we plot the number abundance and the upper  limits of each detected element, Z, as $[{\rm Z/Fe}] = \log{({\rm Z/ Fe})} - \log{({\rm Z/Fe})}_{\sun}$, where the Solar reference abundances are derived from  \citet[][]{asplund09}.

The remarkable  similarity in the average abundance pattern of \lp\ stars is confirmed by a typical dispersion in abundance of order 0.25\,dex, which is of the order  of the uncertainties, for most of the detected elements. We especially note the characteristic $\alpha$ enhancement and moderately super-Solar abundances of the iron-peak elements Cr, Mn, and Ni. One element that is detected in both \jg\ and \jb, Sc, shows the largest mutual scatter ($\approx2$\,dex). Co and Zn, also detected in \jg\ and \jb, have the largest super-Solar abundances among the iron-group elements. Cu was marginally detected in \jg, also with a super-Solar abundance. As noted in Section\,\ref{sec:three-3}, the systematic uncertainties, which we estimated from the analysis of \lp, affect the absolute scaling of trace metals, although they preserve the relative [Z/Fe]. The major effect is seen for O, Ne, and Mg, which are the bulk atmospheric constituents (see next section).

Our revised analysis of \lp\ with respect to \citetalias{raddi18a}  now indicates an abundance, $[{\rm Mn/Fe}] = 0.07$, that  is closer to the Solar reference. Although the uncertainties on the abundances of \lp\  are relatively larger than those estimated for the other  two  stars (cf Section\,\ref{sec:three}), it  appears as the least Mn-rich of the three, with \jg\ and  \jb\ having [Mn/Fe] = 0.43 and  0.63\,dex, respectively. Given the super-Solar trend, we  confirm the evidence in favour of Mn-enhancement for this  group  of stars. Three other elements, Cr-, Co- and Ni-to-Fe confirm a similar Super-solar trend, 
which strengthen the suggested connection with near-M$_{\rm Ch}$ thermonuclear  explosions  \citepalias{raddi18a}. The nucleosynthesis of these elements is known to be enhanced by electron capture that is thought to occur when the conditions for nuclear statistical equilibrium are reached at the high central densities of near-M$_{\rm Ch}$ explosions ($\rho > 2\times10^{9}$\,g\,cm$^{-3}$), which are expected to occur through mass-accretion from a non-degenerate companion, and not by exploding a sub-M$_{\rm Ch}$ white dwarf merger of $\sim 0.8$\,M$_\odot$ \citep{iwamoto99,seitenzahl13a}.

Super-Solar detections of Cu and Zn also agree with the observations of normal stars, for  which the abundance pattern of these elements is proposed to need  a significant  contribution from thermonuclear  supernovae that previously enriched the  interstellar medium with their nucleosyntheis yields \citep{matteucci93}. The close-to-Solar detection of the $p$-nucleus, Sr,  in \jg\ and \jb\ is also interesting, as this heavier element is expected to undergo significant production in the external layers of white dwarfs during thermonuclear explosions \citep{travaglio11}. Finally, we also note that ratios of iron-peak elements (e.g.\ Mn-to-Fe, Ni-to-Fe, and Mn-to-Cr mass fractions) deserve further investigation, as they could be age-estimators  for  the progenitors of \lp\ stars \citep[e.g.\ via element-ratio vs metallicity relations like those employed for the characterisation of supernova nebular remnants;][]{badenes08,yamaguchi15}. 

\subsubsection{Comparison with theoretical yields}
 \label{sec:five-12}
In \citetalias{raddi18a}, we suggested that \lp\  was  unlikely to be the donor star in a binary supernova because its atmospheric composition, enriched with $\alpha$ and iron-peak elements, is incompatible  with the pollution from nucleosynthesis yields of both thermonuclear and core-collapse  supernovae. Considering the scenario  proposed  by \citet{vennes17}, in which \lp\ could be the surviving white dwarf from a SN~Iax explosion, in \citetalias{raddi18a} we compared its atmospheric composition to the bulk composition of theoretical ``bound remnants'' that form in three-dimensional hydrodynamic simulations of pure deflagrations of CO white dwarfs, which were performed  by \citet{kromer13} and  \citet{fink14} to reproduce  the observed  light curves and spectra of SNe\,Iax.  

Although the abundance  pattern of \lp\ followed  the general trend of nuclear yields computed by \citet{fink14}, we noted  two major conflicts  with the  predicted  bulk-composition of bound  remnants, concerning the He-dominated atmosphere and the non-detection of C \citep[the  latter  was  also previously noted by][]{vennes17}. Thanks to a more detailed analysis of \lp\ with the inclusion of {\em HST}/STIS observations and the analysis of \jb\ and  \jg\ (Section\,\ref{sec:three}), we  have now solved the first issue by discarding the He-dominated  atmosphere.

\begin{figure}
\includegraphics[width=\linewidth]{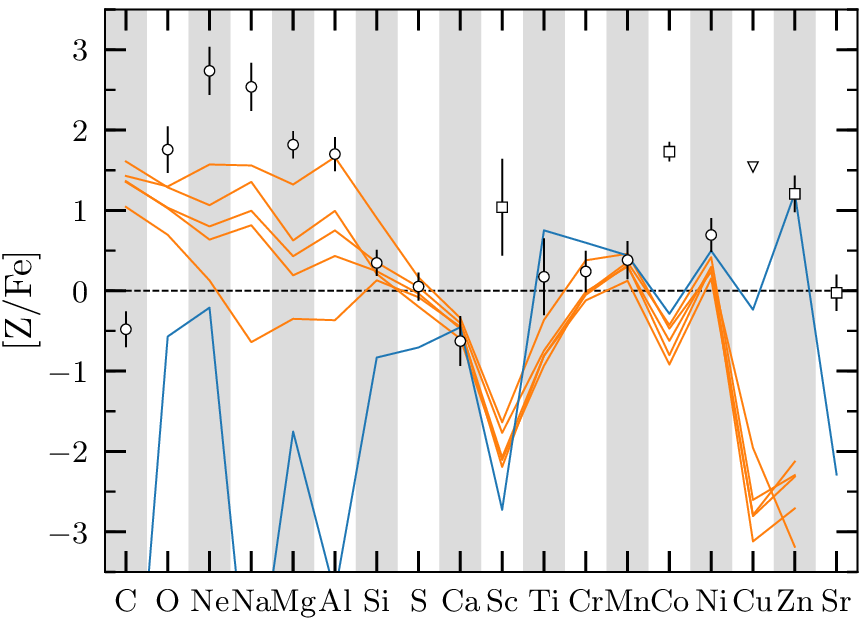}
\caption{The average composition of \lp, \jg, and \jb, scaled  to Solar as in Fig.\,\ref{f:abd}, is compared to the bulk-composition of bound remnants resulting from three-dimensional hydrodynamic simulations of pure deflagrations of  CO \citep{fink14} and ONe white dwarfs \citep[][]{jones19}, plotted as orange and blue curves, respectively. The circles, squares, and triangle indicate whether three, two, and one star, respectively, have been used to compute the average abundances for a given element. The error bars represent the standard deviation of the variance for each element. Upper limits are not taken into account.  \label{f:abdmd}}
\end{figure}

Given the similar abundance pattern of the three \lp\ stars, in Fig.\,\ref{f:abdmd}, we compare their average composition to those of bound  remnants computed by \citet{fink14}. The simulations performed by these authors account for a range  of internal densities of white  dwarfs and explosion intensities,  which lead to a wide variety of light curves and bound remnant  masses (i.e.\ weak explosions produce the  most massive remnants and vice versa).  Although the pure-deflagration models that form low-mass  bound remnants were disfavoured  by \citet{fink14}, because their high energy output does not reproduce the observed SN\,Iax light curves and spectra, we  find them to deliver the best match to the average composition of the \lp\ stars (Fig.\,\ref{f:abdmd}). The best match is obtained  by a $0.1$\,M$_{\sun}$ bound remnant, which corresponds to a progenitor with the highest central density ($5 \times  10^{9}$\,g\,cm$^{-3}$). However,  we note that the observed abundances of the main atmospheric constituents (i.e.\ O, Ne, Mg) still exceed theoretical predictions by 0.5--1.5\,dex. In contrast, C is at least  $1.5$\,dex less abundant than the theoretical curves. The number abundances of trace elements, excluding Sc, Co, Cu, and Zn, are roughly within $1\,\sigma$ of the locus  defined by bound-remnant models. 

Stressing  once  more the  remarkably  similar atmospheric composition of  the  three \lp\ stars, we  suggest that convection, which is predicted  by our models, could play an important role. We  expect the depth of convection zones of \lp\ stars  to  be physically comparable to those of a He-dominated atmospheres of similar \Teff, and we estimated the diffusion timescales of the detected elements to be at least $\sim$10\,Myr. Hence, we speculate that if SN-Iax bound remnant models realistically matched the abundance of Si-burning nucleosynthesis yields detected in  \lp\ stars, either most of the  left-over C might have not  settled  in the  bound remnant or C-burning should be more intense. This interpretation requires testing via  hydrodynamical calculations, which cannot currently resolve the structure of  bound  remnants  and the expansion of  supernova ejecta at the same  time \citep{kromer13}. 

An alternative scenario proposed  by \citet{vennes17} might also find a compromise between low C and high O, Ne, and Mg, abundances, in case the progenitors of \lp\ stars  were hybrid CONe white dwarfs. While the existence of hybrid CONe white  dwarfts is  still debated \citep{denissenkov13,denissenkov15,brooks17a}, one-dimensional computations performed  by \citet{bravo16} showed that thermonuclear explosions of such stars could also form low-mass bound remnants that might represent \lp\ stars. However, we also note that
three-dimensional hydrodynamic simulations of pure deflagrations of hybrid CONe white dwarfs suggest that  bound remnants would be typically more massive \citep[$\simeq1$\,M$_{\sun}$;][]{kromer15}, in contrast with our observations (see next section).

 In  addition to the comparison with bound remnants from CO white dwarf  explosions, we have considered the recent results obtained by \citet{jones16,jones19} for the explosion of mass-accreting ONe  white dwarfs in binary systems. While these binaries are more frequently argued to form neutron stars via the accretion-induced collapse  \citep[][]{schwab15, brooks17b, schwab19},
three-dimensional pure-deflagration simulations performed by \citet{jones16,jones19} have  shown that the nuclear burning  of O and Ne could lift the electron degeneracy and  explode  the  star, leading to what is defined as a thermonuclear electron-capture  supernova (tECSN). Under these  conditions, tECSNe also lead to  the formation of partly burnt bound remnants, instead  of neutron stars. In their work, \citet{jones19} already noted  an overwhelming discrepancy with the abundance of $\alpha$ elements measured for \lp, in contrast  to  a good agreement with the iron-peak elements (Cr, Mn, Ni). Their result is  confirmed via comparison with the average abundance  pattern of the three \lp\ stars in Fig.\,\ref{f:abdmd}. Of the new detected elements, Sc, Co , Cu, and Sr, are discrepant with respect  to the tECSN model. Zn, however, displays a good  match.  For tECSNe  to be viable progenitors of \lp\ stars,  most of the iron-peak elements would need to be confined to the stellar interiors.  This  would be possible either if they were  synthesised in the core of the exploded white dwarf or if they rapidly sunk after the explosion. Hence, atmospheric convection would  still have a role in keeping the surface  abundances as distinct as they are from the bulk composition. Furthermore, in this scenario, we note that \lp\  stars might eventually be the progenitors of the  predicted Fe-core white dwarfs \citep[][]{isern91}.

\begin{figure}
\includegraphics[width=\linewidth]{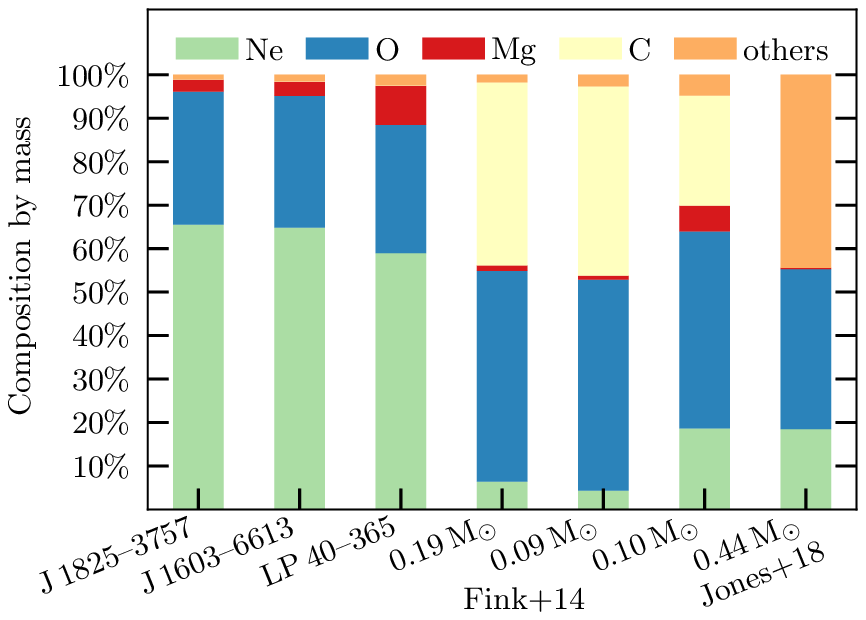}
\caption{The atmospheres  of \lp\ stars compared to the predicted bulk composition of three representative bound-remnants resulting from the pure deflagrations of CO white  dwarfs \citep{fink14} and an ONe white dwarf \citep{jones19}. The 0.19\,M$_{\sun}$ and the 0.09\,M$_{\sun}$ remnants result from the same initial CO white dwarf model, but the former experiences a weaker deflagration with respect to the latter. The 0.09\,M$_{\sun}$ and 0.10\,M$_{\sun}$ bound-remnants, instead, experience deflagrations of equal intensity, but the former has an initial central density that is $\simeq 2$ times smaller than the latter. The ONe white dwarf leading to the 0.44\,$M_{\sun}$ remnant has a similar structure as the CO white dwarfs, with Ne in place of C, and external stratified shells of C, He, and H. \label{f:hist}}
\end{figure}

To wrap up the comparison with theoretical SN-Iax and tECSN bound remnants,  we plotted histogram bars representing the mass fraction of detected elements (Fig.\ref{f:hist}). The remarkable similarity among \lp\ stars is once again evident, with Ne as the most abundant element by mass fraction (59--65~per~cent), followed by O (29--31~per~cent), and Mg (3--9~per~cent).  The remaining elements are just  1-2~per~cent of the total  mass. We note that, considering the systematic uncertainties estimated from the analysis of \lp, O and Ne would change by just a few per cent. However, adopting a 350\,K cooler \Teff, the Mg mass fraction would drop to 2~per~cent and all the trace elements would account  for less than 1~per~cent by mass. The results  from theoretical simulations by \citet{fink14}, instead, contain between 25--40\,per~cent of C and about 45\,per~cent of O. As noted before, the bulk compositions of  theoretical bound remnants are slightly dependent on the explosion intensity and quite sensitive to central density of the accreting white dwarf. On the  other hand, the bound-remnant simulated by \citet{jones19} contains 45\,per~cent of heavy metals, most  of which are  Fe and Ni.

\begin{table*}
    \centering
        \caption{Physical parameters of the \lp\ stars. The $1\sigma$ errors and the 5--95\,per~cent ranges were estimated via Monte Carlo sampling. The apparent fluxes are listed without errors, because they carry very small errors from precise broad-band photometry and we did not include the effect of T$_{\rm eff}$ uncertainties in the error budget. The errors on radius, luminosity, and mass, account for those on parallax, $T_{\rm eff}$, and $\log{g}$.}
    \begin{tabular}{@{}llllllll@{}}
    \hline
Parameters & Symbols & \multicolumn{2}{c}{\lp} & \multicolumn{2}{c}{\jg} & \multicolumn{2}{c}{\jb} \\
& & 16--84\,\% & 5--95\,\% 
& 16--84\,\% & 5--95\,\% 
& 16--84\,\% & 5--95\,\% \\
\hline
Effective Temperature [K] & $T_{\rm eff}$ & \multicolumn{2}{c}{$9800\pm300$} 
& \multicolumn{2}{c}{$10600\pm370$}
& \multicolumn{2}{c}{$12830\pm450$} \\
Surface gravity [cgs] & $\log{g}$ & \multicolumn{2}{c}{$5.50\pm0.30$}
& \multicolumn{2}{c}{$5.34\pm0.20$} 
& \multicolumn{2}{c}{$4.21\pm0.18$} \\

Luminosity [L$_{\sun}$] & $L$ & $0.20\pm0.04$ & 0.14--0.28 
& $0.29^{+0.13}_{-0.10}$& 0.17--0.57
&$8.766^{+1.82}_{-1.57}$& 6.31--11.94\\
Radius [R$_{\sun}$] & $R$ & $0.16 \pm 0.01$ & 0.14--0.17 
& $0.16\pm0.04$& 0.11--0.21
& $0.60\pm0.04$ & 0.54--0.67 \\
Mass [M$_{\sun}$] & $M$ & $0.28^{+0.28}_{-0.14}$ & 0.09--0.87
& $0.20^{+0.16}_{-0.09}$ & 0.07--0.51
& $0.21^{+0.12}_{-0.07}$ & 0.10--0.44\\ 
\hline
    \end{tabular}
    \label{tab:phys}
\end{table*}

Before concluding this section, we note that \citet{zhang19} have recently computed the  evolutionary tracks and  modelled the physical and spectral evolution of post-SN\,Iax bound remnants (named postgenitors by the authors). Taking the \citet{kromer13}  SN\,Iax bound remnant composition as reference, \citet{zhang19} evolved the full stellar structure, including convection and diffusion.
 In contrast with our observations, their simulations, which considered a representative model of 0.6\,M$_{\sun}$ with an envelope accounting for 10~per~cent of the mass, showed that, at late  times after the bound remnants have formed ($\gtrsim 10$\,Myr), the atmosphere is radiative and contains about 70, 30, and 1~per~cent of C, O, and Ne, respectively, while  most of the  heavy elements  have  sunk in to the stellar interiors. The  authors suggested that  similar results were obtained for other lower-mass  models considered. 

 In conclusion, the  interpretation of our results, compared  to  theoretical work,  showed  that, whether \lp\ stars formed through either SNe\,Iax or tECSN explosion, it would be necessary that either C-burning was more intense than what is predicted by models of CO white dwarf deflagrations or, vice-versa, less C, O, and Ne were burned in a ONe  white dwarf. In both cases, we stress that theoretical results for the bulk composition of  bound remnants might not reflect the atmospheric abundances, given that initial segregation  of nucleosynthetic yields,  internal mixing due to gravitational settling, diffusion, and convection should definitely have a strong impact on their internal structure, atmospheric composition, and spectral evolution.

\subsubsection{Masses and radii}
 \label{sec:five-13}

Given the new spectral analysis of \lp\ presented here, we have revised our previous estimates of its  physical parameters. We estimated the radius as  $R = (1 / \varpi) \sqrt{f/F}  = 0.16\pm0.01$\,R$_{\sun}$,  where $f$ and $F$  are the extinction corrected flux and the integrated Eddington flux, respectively, both considered in the Pan-STARRS $g$-band.
The luminosity is  $L = 0.20 \pm 0.04$\,L$_{\sun}$, as determined via 
the Stefan-Boltzmann law.
The radius is 11~per~cent smaller than what estimated in \citetalias{raddi18b},  compensating for the 10~per cent increase in $T_{\rm eff}$ and a more  realistic spectral  modelling. Considering the $\log{g} = 5.5\pm0.3$, we inferred a mass  of  $M = gR^{2}/G = 0.28^{+0.28}_{-0.14}$\,M$_{\sun}$ (24~per~cent smaller than that of \citetalias{raddi18b}), where $G$ is the gravitational constant.

For \jg, we adopted the radius inferred from the spectro-photometric  fit presented in Section\,\ref{sec:three-2}, $R = 0.16\pm0.03$\,R$_{\sun}$, which implies $L = 0.28^{+0.13}_{-0.10}$\,L$_{\sun}$. 
 On the other hand, the smaller error on $\log{g}$ implies a more precisely constrained mass of $0.21^{+0.18}_{-0.09}$\,M$_{\sun}$.

While \lp\ and \jg\ are essentially spectroscopic twins, \jb\ is hotter and intrinsically brighter than the other two stars, and thus  larger. The spectro-photometric fit delivered $R=0.60\pm0.04$\,R$_{\sun}$, which corrsponds to $L = 8.8^{+1.8}_{-1.6}$\,L$_{\sun}$, i.e.\ $30$ times more luminous than the two cooler stars. The inferred mass of \jb, $M = 0.21^{+0.12}_{-0.07}$\,M$_{\sun}$, matches the estimates for \lp\ and \jg. 

The physical parameters of the three stars are listed in Table\,\ref{tab:phys}, which includes both the $1\,\sigma$ uncertainties and the 5--95\,per~cent ranges. 

\begin{figure}
\includegraphics[width=\linewidth]{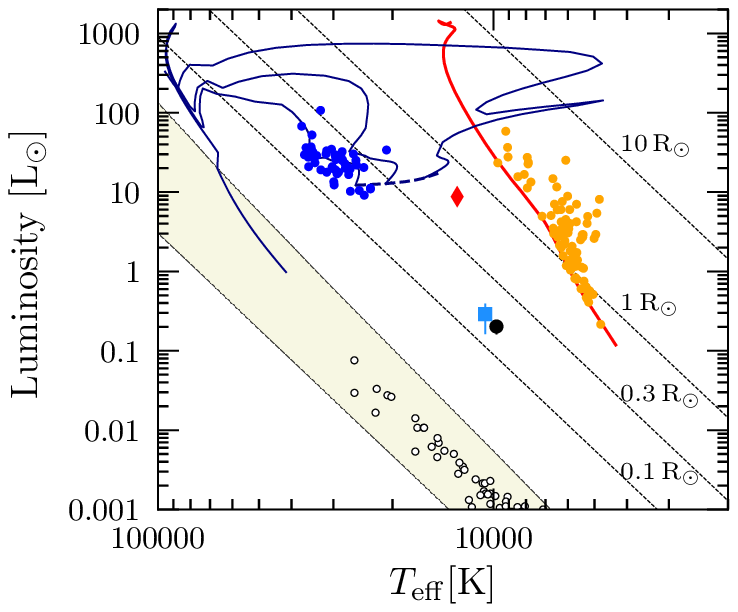}
\caption{Theoretical HR diagram displaying \lp\ (black circle), \jg\ (light-blue square), and \jb\ (red diamond), along with representative samples of hot subdwarfs \citep[blue circles;][]{lisker05}, main-sequence stars \citep[orange circles;][]{boyajian13}, and white dwarfs in the Solar neighbourhood \citep[white circles;][]{giammichele12}. We note that the $1\,\sigma$ errors on the estimated physical parameters are smaller or of the order of the symbol sizes (on the adopted log-log scale). For reference, we draw the canonical-mass white dwarf cooling sequence \citep[beige-coloured strip][]{fontaine01}, the 100 Myr old main-sequence \citep[red curve;][]{choi16}, and the evolutionary tracks for 0.47, 0.48, and 0.50 M$_{\sun}$ hot subdwarfs \citep[blue curves;][]{dorman93} The luminosities for given stellar radii are shown as a function of $T_{\rm eff}$ (dotted curves). \label{f:hrd}}
\end{figure}

 \lp, \jg, and \jb\ are plotted  in the  theoretical HR diagram (Fig.\,\ref{f:hrd}), where  we include  other classes of stars for comparison. While their $T_{\rm eff}$ span just $\sim$2500\,K, their luminosities cover two orders of magnitude (0.1--10\,L$_{\sun}$). This large portion of the HR diagram is not occupied by any long-lived phase of  stellar evolution, but it is occasionally scattered with (pre-) low-mass He-core white dwarfs, which are the products of binary interaction \citep[][]{istrate16}, metal-poor A/F-type stars \citep[][]{pelisoli18}, and short-lived evolutionary phases with stochastic behaviour, like cataclysmic variables or milli-second  pulsar  companions. 
 \jg\ sits very close to \lp, as expected for two spectroscopically and photometrically similar stars. Although \jb\ appears closer to the zero-age extreme horizontal branch and the main sequence (note  the log-log scale), it is $\sim 15\,000$\,K cooler but larger than  equal-luminosity hot subdwarfs. Hence,  it  is at least an order of magnitude less luminous than equal-$T_{\rm eff}$ main sequence stars.
 
 Given their unusual, ``sub-luminous'' location in the HR diagram and their small masses, which are below the typical mass of canonical white dwarfs \citep[e.g.\ 0.621\,M$_{\sun}$;][]{tremblay16}, we speculated that the interiors of \lp\ stars may not be fully degenerate. We measured their average densities as $\overline{\rho} \simeq 340$, 240, and 4.5\,g\,cm$^{-3}$ for \lp, \jg, and \jb, respectively. Hence, we  note that \lp\ and  \jg\ have average densities that are comparable to those of brown dwarfs (or hot subdwarfs), whereas the mean density of \jb\ is similar to that of the Sun. 

Finally, as discussed in \citetalias{raddi18a}, we remark that the radii we measure  for \lp\ stars  are  in agreement with the observed low-rotation rates. We  note that, even if their progenitors were tidally locked in a short period binary \citep[see Section\,\ref{sec:five-5} and][]{fuller12}, we expect that mass loss  and radius expansion caused by the supernova explosion could have both contributed to slow down a potentially fast-rotating star.

\subsubsection{Evolutionary timescales}
 \label{sec:five-14}
 
Given their rarity, the current location of the \lp\ stars in the HR diagram of  Fig.\,\ref{f:hrd}  is likely transitory and it is probably characterised by a relatively fast evolution, which could be shorter than the Galactic crossing time for the unbound stars ($\sim$140\,Myr for \lp; \citetalias{raddi18b}). 

\citet{shen17} modelled the first $\sim$1000~yr of evolution for post-SNe\,Iax bound remnants. In this early phase, the stellar photosphere is initially inflated to $> 1000$\,R$_{\sun}$, the energy output is dominated by strong winds that are powered by the decay of radioactive $^{56}$Ni and $^{56}$Co, and the release of energy takes place over $\simeq$10--100\,yr timescales. More  recent work by \citet{zhang19} covers the  long-term evolution of post-SN~Iax bound remnants, specifically discussing \lp\ as one such star. These authors proposed that post-SN\,Iax bound remnants would initially cool down over relatively short time scales, and  then heat up and brighten during  an at-least 10\,Myr-long timescale until they reach the white dwarf cooling sequence again. This mechanism would be regulated by convection and diffusion of iron-peak elements, which initially would settle in the atmosphere  of the bound remnants for 10,000\,yr during the early phase and would  cause a large opacity. The  long-lived brightening phase would be caused by a sharp decrease of iron-peak element abundances when the bound remnant becomes radiative  and warms up again, after reaching its minimum  luminosity and  coolest temperature that depend on its mass and envelope size.

Despite finding a poor agrement with the atmospheric  composition of \lp, \citet{zhang19} identified a good match between its location in the HR diagram and the evolutionary track computed for a 0.15\,M$_{\sun}$  post-SN\,Iax bound remnant. This result is intriguingly close to the updated masses of \lp\ and \jg. 
Thus, \citet{zhang19} estimated for \lp\ an age of 23\,Myr, which corresponds to the peak  of brightness of the long-lived phase. While \jg\ would have a similar age as \lp, it would seem that \jb\ is older than the other two stars by interpolating between the \citet{zhang19} tracks for 0.15--0.3\,M$_{\sun}$ remnants. Hence, this star might be about to join the white dwarf cooling track over  a relatively sort time.

The presence of iron-peak elements in the atmosphere  of \lp\ stars critically contrasts  with the models  of \citet{zhang19} and it could indicate that the decline of stellar luminosity is characterised by much longer timescales. Hence, the relative ages of the three \lp\ stars could be  inverted, i.e.\ \jb\  and would be the youngest of the three objects. For quantifying this, we estimated the Kelvin-Helmholtz timescales, $\tau_{\rm  KH}$, for the three \lp\ stars that, in absence of internal sources  of energy production, is a good estimator for the evolutionary timescales. Adopting  the formulation $\tau_{\rm KH} = 3GM^{2}/7RL$, which holds  for  a wide range  of objects with internal structure reproducible by a polytrope with  index $n = 3/2$, and using the values  from Table\,\ref{tab:phys}, we obtain $\tau_{\rm KH} \sim 32$, 12, 0.1\,Myr for \lp, \jg, and \jb, respectively. The $\tau_{\rm KH}$ of \lp\ and \jg\  are broadly compatible the evolutionary age estimated by \citet{zhang19}. The $\tau_{\rm KH}$ of \jb\ suggests a much younger age that is, however, larger than the rapid  cooling  phase of the \citet{zhang19} models that is richer in atmospheric iron-peak elements  (100--10,000\,yr). We note that such a short $\tau_{\rm KH}$ is still compatible with the non-association of \jb\ with known diffuse nebulosity caused by young supernova.

\subsubsection{Kinematics and  birth places}
\label{sec:five-5}

The rest frame velocities of \lp\ and \jg\ ($v_{\rm rf} \simeq 800$--850\,\kms) are compatible with the two stars being unbound from the Milky Way. This condition implies that, in absence of a powerful birth kick, both stars took advantage  of a large boost from the Milky Way rotation so that $v^{2}_{\rm rf} = V^{2}_{\rm c} + v^{2}_{\rm ej} + 2V_{\rm c}\,v_{\rm ej}\,\cos{\alpha}\cos{\beta} = 800$--$850$\,\kms, where $\alpha$ and $\beta$ are the angles drawn by the ejection velocity vector, $v_{\rm ej}$, with respect to the vertical direction and the Galactic rotation, respectively. Considering the almost-parallel alignment of the trajectories of \lp\ and \jg\ in the $X-Y$ plane (5--10\,deg with respect to direction of the Galactic rotation in the plane, and 20--26\,deg with respect to the vertical  direction), we found the minimum possible ejection velocities in the range of $v_{\rm ej} \approx 550 $--$600$\,\kms.

In order to achieve the smallest $v_{\rm ej}$ possible, obtainable with the maximum benefit from $V_{\rm c}$, the  supernovae should have exploded at relatively close distances from the Galactic plane and  not to far  away from the  centre \citep[we  note that $V_{\rm c}$ is already $\sim$10\,per~cent slower above $|Z| = 1$\,kpc, and  it  stays  constant up to $R_G \simeq25$\,kpc;][]{williams13,huang16}. This condition is compatible with the prediction for runaway stars ejected by a disc-like distribution of supernova progenitors \citep{kenyon14}. 

Adopting the \citet{zhang19} age estimate of 23\,Myr as the flight  time for both \lp\ and \jg, we could track their  simulated trajectories  back to the fourth and third Galactic  quadrants, respectively. We note that the  $R_G \sim 25$\,kpc estimated by \citet{zhang19} does not account for the past trajectory of \lp, and it is rather equivalent to the radial distance that  would  be  covered  in  23\,Myr at 850\,\kms. Hence, properly tracing  the Galactic  orbits, we find that \lp\ and \jg\ would have been ejected from possible birth sites at  $R_{G} \approx 15$\,kpc  and $Z \approx -1.2$\,kpc, and  $R_{G} \approx 17$\,kpc  and $Z \approx +3.4$\,kpc, respectively. These coordinates are roughly compatible  with birth sites  within the thick disc. Furthermore, we note the Galactocentric radii of the identified birth sites are close to the empirical determination for the edge  of the Galactic disc, i.e\ where most of the star formation is seen to take  place  \citep[within a $R_{\rm G} = 13$--16\,kpc cut-off, as  observed from photometric  surveys of the Milky Way][]{sale10,minniti11}. Thus, this result agrees with SN\,Iax observations, which are mostly seen in association with regions  of recent or ongoing star formation \citep{foley13,jha17,lyman18}. On the other hand, if we wanted to impose birth sites that are  nearer to the plane, say $|Z|=1$, to find a better match with the minimum $v_{\rm ej} = 600$\,\kms,  we would  obtain $R_{G} = 12$,~8\,kpc  and $\tau_{\rm fl}  = 19$,~7.5\,Myr for \lp\ and \jg, respectively.

We previously noted that \jb\ is bound to the Milky Way. The fact that it moves in the opposite direction with respect to the Galactic rotation and it follows a fairly eccentric orbit strongly supports the supernova-ejection mechanism. Observing that the direction of motion of  this star has an equally small inclination with respect to the Galactic plane (a few degrees) as the other two, and it is a $\sim$165\,deg with respect to the Galactic rotation, we also estimated a minimum ejection velocity of $v_{\rm ej} \simeq 600$\,\kms. Finally, we note that if \jb\ is as young as $\tau_{\rm KH}\sim 0.1$\,Myr, it would have formed  within the thin disc. 

\subsubsection{Binary progenitors}
\label{sec:five-16}

Given that we find similar $v_{\rm ej}$  for the three \lp\ stars, it is reasonable to think they formed from the disruption of similar binary progenitors.  In the ideal case  of instantaneous mass loss and no interactions between accretor and donor stars \citep{hills83}, $v_{\rm ej} \sim 550$--$600$\,\kms\ are lower  limits on the  orbital velocity at the moment of explosion. This condition could be met if the progenitors of \lp\ stars were white dwarfs of  $M \geq 1.2$\,M$_{\sun}$, which were orbited by $ 0.8$--1.3\,M$_{\sun}$ He-burning donor stars in $\simeq 1$\,hr  \citepalias{raddi18b}.  This result is compatible with a single-degenerate scenario proposed for SN\,Iax \citep[including progenitors with CO and hybrid CONe  cores;][]{wang09b,wang13,wang14}, and may also be representative of accreting ONe white  dwarfs that  lead  to tECSNe. We note that kicks due to asymmetric  explosions and interactions with the both the  donor star and supernova ejecta  are expected  to take place \citep{marietta00, jordan12, pan12, liu13}, adding complexity  to this interpretation. 

The rest frame velocities of the unbound \lp\ stars contrasts with the  much larger $v_{\rm rf}\gtrsim 1000$\,\kms\ of the other known hyper-runaway stars, i.e.\ the D$^{6}$ stars \citep[proposed as the former donor white dwarfs in double-degenerate supernova progenitors;][]{shen18b} and the hot subdwarf star US\,708 \citep[proposed to be the former low-mass  He subdwarf donor in a single-degenerate sub-M$_{\rm Ch}$  supernova;][]{geier15}. Hence, the kinematics of  \lp\ stars confirms their origin from a distinct class of binary progenitors, also exploding as thermonuclear supernovae, but possessing smaller orbital velocities than the D$^{6}$ and US\,708 progenitors.

\subsection{Prospects for end-of-mission {\em Gaia} detections}
 \label{sec:five-17}

\citet{zhang19} have estimated that four post-SN\,Iax stars would be detectable within 2\,kpc from the Sun. In contrast, \citet{shen18b} estimated the number of D$^{6}$ stars in {\em Gaia} as 14--40  within 1\,kpc. The results obtained by these authors are  comparable, considering that empirical SN\,Iax rates are typically estimated to occur as 1--30~per~cent of the SN\,Ia rates \citep[][]{foley13,li11b,graur17, jha17}. Although \citet{zhang19} considered \lp\ stars as good candidates for the post-SN\,Iax  bound  remnants, they assumed  these runaways to have $v_{\rm  rf}  \sim 1000$\,\kms. Such large velocities have only been measured for D$^{6}$ stars and the hot subdwarf US\,708. Given the smaller $v_{\rm rf}$ of \lp\ stars, likely due to their different binary progenitors, we re-estimated their space density. 

Assuming the progenitors of \lp\ stars have a space density reflecting that of the Galactic thin disc,  we  distributed them according to an exponentially decaying profile (scale-height and scale-length of 0.3 and 3\,kpc, respectively). We scaled the formation rate of \lp\ stars to 10~per~cent of the Galactic SN\,Ia rate of $10^{-13}$~yr$^{-1}$\,M$^{-1}_{\sun}$ \citep[][]{li11c}, as adopted by \citet{shen18b}. This rate is compatible with the average of other estimates from the literature, which include a mix of progenitors from both the single- and double-degenerate channels \citep[i.e.\ SNe\,Ia with a variety of delay times;][]{scannapieco05,mannucci05,mannucci06,aubourg08,ruiter09,maoz11}. Hence, considering a Galactic disc mass of $\simeq5\times10^{10}$\,M$_{\sun}$ \citep{bland-hawthorn16}, we created 50,000 objects with birth-times that were uniformly distributed over a time interval of 100\,Myr, $t_{0} = \mathcal{U}(0,100)$. We assigned Gaussian-distributed initial velocities, $v_{\rm ej} = \mathcal{N}(600, 25)$, which we modelled on those we have inferred for the observed \lp. Finally, we used {\sc galpy} to estimate the kinematic parameters after a flight time  of $\tau_{\rm fl} = (100 - t_{0})$\,Myr. 

Running ten simulations to randomise the initial conditions, we counted $78\pm 8$ objects within 2\,kpc. Such an estimate included all objects with a $\tau_{\rm fl} \leq 100$\,Myr. However, we know that a large fraction of them might not be detectable for several reasons, one of which is the long-term evolution of their absolute magnitudes. If we  considered that \lp\ stars evolve as  proposed by \citet{zhang19}, i.e.\ they are 10--1000 times  fainter than \lp\ in their first $\sim10$\,Myr of life, we would  expect that  only those older than  $20$\,Myr (i.e. $63\pm6$) are detectable by {\em Gaia} with $G<18$ (assuming the absolute magnitude of \lp, $M_{G} = 6.5$, as a lower  limit). Hence, we could expect to detect just  1--3 intrinsically fainter, younger \lp\  stars in a smaller volume (e.g.\ $d < 1$\,kpc) down to $G  = 19$. 

The estimate of 63 stars might be optimistic, given that so far we have detected just three \lp\ stars  within 2\,kpc,  suggesting that the timescales  proposed  by \citet{zhang19} are too short. Following this interpretation, all the observed \lp\ stars  are still relatively young, having ages of the same order  of magnitude  or shorter than the $\tau_{\rm KH} \sim 32$\,Myr measured for \lp. In this case, we would expect $22\pm4$ \lp\ stars. Although this estimate is still  too large in comparison with the observed numbers, we note  a better agreement with the predictions by \citet{shen18b}, if we scale our estimate to the number of D$^{6}$ stars within a smaller volume  with a 1\,kpc radius.

Given their relatively  low $v_{\rm ej}$, most of the \lp\ stars could  stay bound to the Milky Way,  as we find that just $\simeq$20~per~cent of the simulated objects have $v_{\rm rf} > v_{\rm esc}$. The fraction of stars achieving $v_{\rm rf} > 1.3\,v_{\rm esc}$ is just 4~per~cent of the total. Typical proper  motions  of \lp\ stars would also be relatively small, with an average value of $\mu = 49^{+33}_{-20}$\,mas, indicating that \lp\ and \jb\ could be the rarest outliers in the $\mu$ distribution. 

While we stress that these simulations are affected by several unknowns such as the confirmation of a cooling law and the variety of supernova rates, we note that they are sufficiently straightforward to motivate further spectroscopic follow-up of \lp\ star candidates. We  also note that the binary-supernova mechanism implies the existence of ejected  donor stars, which may have similar
kinematics, but different compositions.  

Finally, we note that the results of this section can also be extended to the ONe white dwarfs that may explode as tECSNe, for which \citet{jones19} estimated a birth rate of $\sim 10^{-14}$\,yr$^{-1}$\,M$_{\sun}^{-1}$, which is of the same order of magnitude  of that adopted for SN-Iax bound remnants. Thus, tECSNe might lead to the formation of $\sim20$ detectable \lp\ stars within 2\,kpc from the Sun.

\section{Summary and conclusions}
 \label{sec:six}
 
Through detailed spectroscopic, photometric, and kinematic analysis, we have gained further evidence in favour of the idea initially advanced by \citet{vennes17} and then reinforced in \citetalias{raddi18a} and \citetalias{raddi18b} that the high-velocity star \lp\ could be a partly burnt white dwarf accretor that survived to a peculiar class  of thermonuclear supernovae occurring in binary systems. Given the striking spectroscopic and kinematic similarity we found among \lp\ and three new objects (\jg, \jb, \jc), we propose this star as the namesake of its own spectral class  that is defined both by spectroscopic, physical, and kinematic properties:

\begin{enumerate}
\item  Ne-dominated atmospheres, with O and Mg as the second- and
third-most-abundant elements, respectively.
\item Broadly homogeneous composition of detected trace elements (C, Na, Al, Si, S, Ca, Sc, Ti, Cr, Mn, Fe, Co, Ni, Cu, Zn, Sr).
\item Low mass, below the canonical white dwarf mass.  
\item Ejection velocity in the range of $550$--$600$\,\kms.
\end{enumerate}

The atmospheric composition of \lp\ stars is a strong indication of partial C-, O-, and Si-burning, likely connected to thermonuclear explosions that did not entirely disrupt their progenitors. Of significant note is the near-identical spectroscopic properties of both \lp\ and \jg, which strongly suggests a similar formation mechanism. 
We found some similarities with the simulations of pure deflagrations of CO and ONe white dwarfs, which could be representative of SNe\,Iax \citep{fink14} and thermonuclear electron-capture supernovae \citep{jones19}, respectively. We noted that the models  do not reproduce the main atmospheric  components  (O, Ne, Mg), perhaps because the  bulk composition of bound remnants is not represented by the photosphere of these stars. Promising simulations by \citet{zhang19} describe,  for the first time, the evolution of post-SN\,Iax stars. However, the temporal  evolution does not well match those of the \lp\ stars.  New theoretical evolutionary models are required,  which better account for the initial stratification of nucleosynthetic yields  within bound remnants, internal  mixing, convection, and diffusion, and also including the evolution of hybrid CONe white dwarfs.

The physical properties of \lp\  stars are precisely constrained by our spectral analysis and {\em Gaia} parallaxes, indicating a relatively small range of masses (median masses between 0.20--0.28\,M$_{\sun}$), which indicates they currently are (not fully degenerate) low-mass objects. The radii of these stars were likely inflated due to the large energy accumulated during and after the supernova explosion, growing to 0.16--0.60\,R$_{\sun}$, roughly an order of magnitude larger than typical white dwarf radii. Such a large range is compatible with the stars being caught at different evolutionary stages. While the \citet{zhang19} models predicted the \lp\  stars could be older than $\sim23$\,Myr, we speculated that \lp\ stars could be still relatively unevolved and thus younger than the Kelvin-Helmholtz timescale of \lp\ ($\sim 32$\,Myr), due  to the large abundance of heavy metals in their atmospheres.

The kinematics of \lp\ stars are also  distinctive, indicative of their formation mechanism. We found that  they can be either bound or unbound to the Milky Way, depending on whether the stars were ejected in the direction of Galactic rotation (as with \lp\ and \jg) or in the opposite direction (like \jb). In order  to obtain the observed kinematics, confined to small $Z$ coordinates, the progenitors of \lp\ stars likely resided in the Galactic disc, further constraining their ages to no more than a few 10\,Myr. The ejection velocities required to reproduce the rest-frame velocity of the three \lp\ stars are in the range  of 550--600\,\kms, consistent with ejection from a relatively compact binary systems containing a the near-M$_{\rm  Ch}$  white  dwarf  progenitor and a He-burning donor star of roughly 0.8--1.3\,M$_{\sun}$ orbiting with a $\sim 1$\,hr period. This result is at least  compatible with one class of thermonuclear supernova progenitors, which are invoked to explain short-delay-time supernovae and subluminous classes  like SNe\,Iax \citep{wang09b, wang09c, wang09a, wang13}. Furthermore, we noted that the kinematic evidence characterising \lp\ stars clearly distinguishes them from other groups of hyper-runaways, i.e.\ the D$^{6}$ stars \citep{shen18b} and the hot-subdwarf US\,708 \citep{geier15}, which require even more compact progenitors to reach their rest frame velocities of $\gtrapprox  1000$\,\kms.

We estimate that of the order 22 stars could be detectable by the end of the {\em Gaia} mission if \lp\ stars are  post-SN\,Iax stars, imposing the boundary conditions implied  by the ejection velocities  and flight times determined from our analysis. Although the results of our simulations were susceptible to large variations, e.g.\ the uncertainty of supernova rates and evolutionary timescales of  \lp\ stars, we stress they are compatible with the observed numbers and scale as 10~per~cent of of predicted D$^{6}$ stars \citep{shen18b}. We also noted that the results of our predictions for future {\em Gaia} detection are compatible  with the possibility that \lp\ stars  could be bound  remnants  surviving to tECSNe.

The key message from our simulation of the population of \lp\ stars within 2\,kpc of the Sun is that most  of them are likely gravitationally bound to the Milky Way and will possess relatively small proper motions. The confirmed \lp\ stars may not represent the broader population yet to be discovered, suggesting that evolutionary ``missing links''  as  well as their ejected donor stars are yet to be found. It is possible that more evolved \lp\ stars would show different atmospheric compositions, if their heavy elements sink deeper into the atmosphere as the stars evolve with changing temperatures and radii. We anticipate interesting results from more advanced theoretical modelling, expanding the results  presented by \citet{fink14}, \citet{kromer15}, and \citet{jones19}  for  the composition of bound remnants and \citet{zhang19} for  the late cooling of SN\,Iax bound remnants.

Finally, we remark that the \lp\ stars that remain bound to the Milky Way could revert back to the white dwarf cooling sequence, likely within a few 100\,Myr, still displaying lower $\log{g}$ than canonical white  dwarfs due to their smaller masses. These white dwarfs should have stratified atmospheres, containing mostly O, Ne, and Mg. Although not fully investigated, peculiar O-rich white dwarfs like \dox\ \citep{kepler16} -- as well as one of our SDSS candidates, \jd\ -- could represent or be linked to one of  the final evolutionary stages of the newly discovered class of \lp\ stars.

\section*{Acknowledgements}

We thank S.~O.~Kepler, R.~Fisher, A.~Rebassa-Mansergas, S.~Toonen, M.~Zhang, and C.~Wheeler for their helpful feedback during EUROWD21 at the University of Texas at Austin,  
A.~Irrgang for discussions on Galactic trajectories, and S.~Jones for kindly providing his bound remnant  compositions in machine-readable  format.

RR and UH acknowledge funding by the German Science foundation (DFG) through grants HE1356/71-1 and IR190/1-1. Support for JJH was provided by NASA through Hubble Fellowship grant \mbox{\#HST-HF2-51357.001-A}, awarded by the Space Telescope Science Institute, which is operated by the Association of Universities for Research in Astronomy, Incorporated, under NASA contract NAS5-26555. IP acknowledges support from DFG under grant GE 2056-12-1. JS acknowledges support from the Packard Foundation. The research leading to these results has received funding from the European Research Council under the European Union's Horizon 2020 research and innovation programme no. 677706 (WD3D) and the European Union's Seventh Framework Programme (FP/2007-2013) / ERC Grant Agreement no. 320964 (WDTracer). OFT and BTG were supported by the UK STFC grant ST/P000495. 

This work has made use of data from the European Space Agency (ESA) mission {\it Gaia} (\url{https://www.cosmos.esa.int/gaia}), processed by the {\it Gaia} Data Processing and Analysis Consortium (DPAC, \url{https://www.cosmos.esa.int/web/gaia/dpac/consortium}). Funding for the DPAC has been provided by national institutions, in particular the institutions participating in the {\it Gaia} Multilateral Agreement. Based on observations made with the NASA/ESA HST, obtained at the Space Telescope Science Institute, which is operated by the Association of Universities for Research in Astronomy, Inc., under NASA contract NAS 5-26555. These observations are associated with program 15431. Based on observations made with ESO Telescopes at the La Silla Paranal Observatory under programme ID 093.D-0431, 097.D$-$1029, and 0101.C$-$0646. Based on observations obtained at the Southern Astrophysical Research (SOAR) telescope, which is a joint project of the Minist\'{e}rio da Ci\^{e}ncia, Tecnologia, Inova\c{c}\~{o}es e Comunica\c{c}\~{o}es (MCTIC) do Brasil, the U.S. National Optical Astronomy Observatory (NOAO), the University of North Carolina at Chapel Hill (UNC), and Michigan State University (MSU). Also based on observations collected at Copernico telescope (Asiago, Italy) of the INAF -- Osservatorio Astronomico di
Padova. The William Hershel Telescope and its service programme are operated on the island of La Palma by the Isaac Newton Group of Telescopes in the Spanish Observatorio del Roque de los Muchachos of the Instituto de Astrof\'isica de Canarias (prog. ID: W/2017A/25, W/2017A/30, andSW2017a12).

Funding for the Sloan Digital Sky Survey IV has been provided by the Alfred P. Sloan Foundation, the U.S. Department of Energy Office of Science, and the Participating Institutions. SDSS-IV acknowledges support and resources from the Center for High-Performance Computing at the University of Utah. The SDSS web site is www.sdss.org.

This research made use of Astropy, a community-developed core Python package for Astronomy \citep{astropy18}.




\bibliographystyle{mnras}
\bibliography{lp40stars}





\bsp	
\label{lastpage}
\end{document}